\documentclass[twocolumn]{aastex63}

\usepackage{amsmath}
\usepackage[utf8]{inputenc}

\begin{document}

\title{A Multi-Data Approach to Open Clusters: Roslund 3 and Ruprecht 174 in CCD UBV and Gaia DR3 Context}

\author[0009-0004-5726-3749]{H\"ulya Karag\"oz}
\affiliation{Istanbul University, Institute of Graduate Studies in Science, Programme of Astronomy and Space Sciences, 34116, Istanbul, Turkey}
\email{hulyaercay5@gmail.com}  

\author[0000-0002-5657-6194]{Talar Yontan} 
\affiliation{Istanbul University, Faculty of Science, Department of Astronomy and Space Sciences, 34119, Beyaz\i t, Istanbul, Turkey}

\author[0000-0003-3510-1509]{Sel\c{c}uk Bilir} 
\affiliation{Istanbul University, Faculty of Science, Department of Astronomy and Space Sciences, 34119, Beyaz\i t, Istanbul, Turkey}

\author[0000-0002-0435-4493]{Olcay Plevne} 
\affiliation{Istanbul University, Faculty of Science, Department of Astronomy and Space Sciences, 34119, Beyaz\i t, Istanbul, Turkey}

\author[0000-0002-0688-1983]{Tansel Ak} 
\affiliation{Istanbul University, Faculty of Science, Department of Astronomy and Space Sciences, 34119, Beyaz\i t, Istanbul, Turkey}
\affiliation{Istanbul University, Observatory Research and Application Center, Istanbul University, 34119, Istanbul, Turkey}

\author[0000-0002-0912-6019]{Serap Ak} 
\affiliation{Istanbul University, Faculty of Science, Department of Astronomy and Space Sciences, 34119, Beyaz\i t, Istanbul, Turkey}

\author[0000-0003-2575-9892]{Remziye Canbay} 
\affiliation{Istanbul University, Faculty of Science, Department of Astronomy and Space Sciences, 34119, Beyaz\i t, Istanbul, Turkey}

\author[0000-0002-5657-6194]{Timothy Banks} 
\affiliation{Nielsen, 675 6th Ave, New York, NY 10011, USA}
\affiliation{Department of Physical Science and Engineering, Harper College, 1200 W Algonquin Rd, IL 60067, USA}


\begin{abstract}
A detailed analysis of the structural, astrophysical, kinematic, and dynamical properties of the open clusters Roslund~3  and Ruprecht~174 is carried out using CCD {\it UBV} photometry in conjunction with astrometric and photometric data from {\it Gaia} DR3. Membership probabilities were computed via the {\sc UPMASK} algorithm applied to {\it Gaia} proper motions and trigonometric parallaxes, leading to the identification of 198 likely members for Roslund~3 and 397 for Ruprecht~174. Astrophysical parameters were derived using both the classical approach, where parameters are independently determined, and a MCMC technique, which estimates them simultaneously. The agreement between the results from both methods confirms their reliability and highlights the robustness of the classical method. The reddening values were determined as $E(B-V)=0.410\pm 0.046$~mag for Roslund~3 and $E(B-V)=0.615\pm 0.042$~mag for Ruprecht~174. The estimated distances are $d=1687 \pm 121$~pc for Roslund~3 and $d=2385 \pm 163$~pc for Ruprecht~174.  Both clusters exhibit metallicities close to the solar value, with [Fe/H] = $0.030 \pm 0.065$~dex for Roslund~3 and [Fe/H] = $0.041 \pm 0.064$~dex for Ruprecht~174. The corresponding ages were found to be $\tau=60\pm 6$ and $\tau=520\pm 50$~Myr, respectively. The present-day mass function slopes were found to be $1.18 \pm 0.13$ for Roslund~3 and $1.53 \pm 0.30$ for Ruprecht~174, consistent with the canonical Salpeter value within uncertainties. Galactic orbital analyses indicate that both clusters are thin-disk members confined within the Solar circle. Additionally, relaxation times and spatial distributions of stars suggest that both clusters have reached dynamical relaxation and exhibit clear signs of mass segregation.
\end{abstract}





\section{Introduction} 
Open clusters (OCs) are considered crucial astrophysical laboratories for studying star formation and evolution, and the structural dynamics of the Galactic disk. These systems form through the gravitational collapse of molecular clouds. They host stars with similar physical and chemical properties \citep{Lada03, Maurya20}. Typically young and metal-rich, OCs are predominantly located near the Galactic plane, playing a significant role in understanding the evolutionary processes of the Milky Way \citep{McKee07}.

The determination of fundamental astrophysical parameters for these clusters relies on photometric, spectroscopic, and astrometric methods \citep{Dias21}. Properties such as age, distance, and metallicity, which are often challenging to obtain for individual stars, can be more reliably derived through the collective analysis of cluster members. Recent high-precision astrometric studies, such as the {\it Gaia} mission \citep{Gaia_mission}, have significantly enhanced the ability to study the motions of cluster members, thereby improving our understanding of Galactic structure and dynamic processes \citep{Tarricq21, Rangwal24, Cinar24}.

Due to dynamic processes, open clusters evolve over time, typically dispersing within a few hundred million years \citep{Portegies-Zwart10}. Stellar feedback, tidal forces, and interactions with giant molecular clouds are key mechanisms accelerating this process. Despite the transient nature of clusters, their dissolution contributes to the stellar population of the Galactic disk, influencing stellar migration and the long-term evolution of Galactic structure \citep{Piskunov06, Bastian18}. Furthermore, internal dynamical processes such as mass segregation and binary star interactions play a critical role in shaping their evolutionary paths.

Moreover, OCs are invaluable tools for tracing the large-scale structure of the Galaxy. By combining kinematic data with astrophysical properties, researchers have been able to map critical structural features of the Milky Way, including its spiral arms \citep{Castro-Ginard21, Hao21}, warp of the disc \citep{Vazquez10, Hidayat19}, and radial metallicity gradients \citep{Spina22, Joshi24}. However, determining these fundamental parameters remains challenging due to the complex relationships between age, reddening, and metallicity \citep{Yontan15, Ak16}. To mitigate the effects of parameter degeneracy and achieve more reliable results, it is essential to apply independent analytical techniques that account for the specific physical conditions of each cluster.

As part of the systematic OC surveys that we began almost 15 years ago \citep[c.f.][]{Bilir10, Yontan15, Bostanci15, Yontan23a}, we integrate {\it UBV} photometric observations with the high-precision astrometric and radial velocity data obtained from the {\it Gaia} mission \citep{Gaia_mission}. In line with the vision of the present study, the utilization of {\it UBV} photometric data provides an opportunity to independently determine the reddening and metallicity of clusters, which in turn allows for a more accurate and precise estimation of their distances and ages. This methodology facilitates a comprehensive examination of the structural, astrophysical, kinematical, and dynamical orbital characteristics of OCs, thereby enabling us to explore their formation sites.

We focus on Roslund 3 and Ruprecht 174 to enhance our understanding of relatively under-explored OCs within the Milky Way in this study. Due to their proximity to the Galactic plane, these clusters are significantly affected by the presence of field stars, necessitating the use of statistical techniques to minimize the contamination caused by these external sources. This work continues a systematic research initiative aimed at examining lesser-studied OCs, thereby contributing to a more comprehensive understanding of the Galactic structure and its evolutionary processes \citep{Yontan15, Yontan19, Yontan21, Yontan22, Yontan23b, Banks20, Akbulut21, Koc22, Cakmak24}. By utilizing CCD {\it UBV} photometric data in conjunction with the most current datasets provided by the {\it Gaia} mission, we aim to derive the fundamental parameters of Roslund 3 and Ruprecht 174 with enhanced accuracy.

Roslund 3 (C1956+203 in the IAU nomeclature), at $\alpha=19^{\rm h} 58^{\rm m} 46^{\rm s}.80$, $\delta=+20^{\rm d} 30^{\rm m} 32^{\rm s}.5$, $l= 58^{\rm o}.842$, $b=-4^{\rm o}.684$ (J2000), Trumpler class: IV1p, was described by \cite{Turner93} as a `sparse, unstudied, open cluster'. Roslund 3 was discovered by \cite{Roslund_1960}, who noted it was of diameter 10 arcminutes and composed of mid and late B main-sequence stars. Later studies were mainly surveys of OCs, using data sources such as the Tycho-2 \citep{Hog_2000}, DAML02 \citep{Dias_2002}, UCAC4 \citep{Zacharias_2013}, and PPMXL \citep{Roeser10} catalogs \citep{Loktin03, Kharchenko05, Kharchenko13}. Table~\ref{tab:literature} presents key parameter estimates from subsequent research. While agreement between studies is closer for the more recent papers, there are still substantially different estimates, such as the radial velocity and iron abundances of \citep{Soubiran18, Dias21}. There is also a wide scatter in the cluster age estimates from 363 to 41 Myr, with no clear agreement in even recent papers. 

Ruprecht 174 (C2041+368), at $\alpha=20^{\rm h} 43^{\rm m} 25^{\rm s}.44$, $\delta=+37^{\rm d} 01^{\rm m} 51^{\rm s}.7$, $l= 78^{\rm o}.012$, $b=-3^{\rm o}.377$ (J2000), was first discussed by \cite{Ruprecht66}, who used the \cite{Trumpler30} class II2p to describe it (detached with little central concentration, moderate range in brightness, less than 50 stars apparent in the cluster). \cite{Bonatto10} derived fundamental, photometric, and structural parameters for 15 Ruprecht clusters, including Ruprecht 174, based on 2MASS photometry \citep{Skrutzie16}. Subsequent papers also made use of catalogs, as for Roslund 3. Key estimates are summarized in Table~\ref{tab:literature}. These, too, show a scatter in parameter estimates even between the most recent studies. 


\begin{table*}
    \centering
    \setlength{\tabcolsep}{3pt}
    \renewcommand{\arraystretch}{0.8}
    \small
    \caption{The fundamental parameters of Roslund 3 and Ruprecht 174, gathered from the literature,} include reddening for $E(B-V)$, distance $d$, iron abundance [Fe/H], age $\tau$, proper-motion components $\langle\mu_{\alpha}\cos\delta\rangle$ and $\langle\mu_{\delta}\rangle$, radial velocity $\langle V_{\rm R}\rangle$, the corresponding references, as well.
    \begin{tabular}{cccccccc}
    \hline
    \hline
    \multicolumn{8}{c}{Roslund 3}\\
        \hline
        \hline
$E(B-V)$                    & $d$                   & [Fe/H]            & $\tau$               &  $\langle\mu_{\alpha}\cos\delta\rangle$   &$\langle\mu_{\delta}\rangle$ & $\langle V_{\rm R}\rangle$       & Ref \\
(mag)                       & (pc)                  & (dex)             & (Myr)             & (mas yr$^{-1}$)       & (mas yr$^{-1}$)   & (km s$^{-1})$                                   &    \\
    \hline
0.25                        & $2290\pm50$           & ---               & 90                & ---                   & ---               & $-2.6\pm1.1$                                    & (1) \\
---                         &  ---                  & ---               &  ---              & $-2.18\pm0.25$        & $-6.28\pm0.32$    & ---                                             & (2) \\
0.34                        & 1515                  & ---               & 41                & $-0.70\pm0.35$        & $-3.98\pm0.46$    & ---                                             & (3) \\
---                         & $1467\pm293$          & ---               & 109               & $-1.61\pm0.47$        & $-5.28\pm0.47$    & $-4.7\pm3.0$                                     & (4) \\  
0.34                        & 1515                  & ---               & 41                &  ---                  &        ---        & ---                                            & (5) \\  
---                             & $1467\pm293$    & ---               & $109\pm27$        & $-1.61\pm0.47$        & $-5.28\pm0.47$    & $-4.7\pm3.0$                                    & (6) \\  
0.323                       & 1582                  & ---               & 50                & $-1.72\pm0.33$        & $-5.35\pm0.33$    & $-2.60\pm0.32$                                   & (7) \\
---                         & ---                   & ---               & ---               & $-0.98\pm0.66$        & $-4.74\pm0.10$    & ---                                             & (8) \\

$0.340\pm0.032$             & 1522                  & ---                & $106\pm51$       &$-2.385\pm0.191$       & $-8.765\pm0.225$  &$-2.60\pm0.32$                                   & (9)  \\
---                         & ---                   &  ---               & ---              & $-2.440\pm1.670$        & $-4.750\pm1.650$    & ---                                             & (10) \\
---                         &  ---                  &  ---               &  ---             & ---                   &  ---              & $-21.67\pm1.10$                                  & (11) \\
---                         & $1629^{+317}_{-229}$  & ---                & $69\pm10$        &$-2.963\pm0.068$       & $-6.163\pm0.101$  & $-2.26\pm0.21$                                   & (12) \\
---                         & $1727\pm86$           & ---                & $363\pm22$       &$-0.379\pm0.142$       & $-5.088\pm0.118$ & ---                                             & (13) \\
---                         & $1080\pm13$           & 0.00                & $44\pm3$         &---                    & ---               & ---                                             & (14) \\
0.248                       & 1616                  & ---                & 54               &$-0.386\pm0.086$       & $-5.108\pm0.101$  &---                                              & (15) \\
---                         & $1629^{+229}_{-317}$  & ---                & ---              &$-0.386\pm0.010$       & $-5.108\pm0.011$  & ---                                             & (16) \\
$0.376\pm0.010$             & $1583\pm43$           & $0.197\pm0.095$    & $109\pm39$       &$-0.376\pm0.010$       & $-5.107\pm0.012$  & ---                                             & (17) \\
$0.371\pm0.019$             & $1617\pm28$           & $0.112\pm0.080$     & $105\pm11$       & ---                   & ---               & ---                                             & (18) \\
0.312                       &1652                   &  ---               & 75               &$-0.405\pm0.116$       & $-5.153\pm0.119$  & $-2.93\pm10.47$                                 & (19) \\

  \hline
  \hline
    \multicolumn{8}{c}{Ruprecht 174}\\
        \hline
        \hline
$E(B-V)$ & $d$ & [Fe/H] & $\tau$ &  $\langle\mu_{\alpha}\cos\delta\rangle$ &  $\langle\mu_{\delta}\rangle$ & $\langle V_{\rm R}\rangle$ & Ref \\
    (mag) &  (pc)  & (dex) & (Myr) & (mas yr$^{-1}$) & (mas yr$^{-1}$) & (km s$^{-1}$) & \\
    \hline
$0.32\pm0.06$               & $2110\pm200$           & ---                & $800\pm100$      & ---                    &  ---             & ---                                             & (20) \\
0.321                       & 1817                   & ---                & 794              &$-2.30\pm0.34$          &$-2.66\pm0.34$    & ---                                             & (07) \\
---                         & ---                    & ---                & ---              & $-2.73\pm3.61$         & $-4.49\pm2.15$   & ---                                             & (08)\\ 
0.435                       & 2599                   &  ---               & 676              & $-2.73\pm0.23$       & $-4.49\pm0.14$ & ---                                             & (09) \\
---                         & ---                    & ---                & ---              & $-3.20\pm1.22$         & $-5.33\pm0.91$   & ---                                             & (10)\\
---                         & ---                    & ---                & ---              & ---                    & ---              & $-12.89\pm4.49$                                 & (11) \\
---                         & $2328^{+429}_{-735}$   & ---                & ---              &$-3.154\pm0.084$        & $-4.675\pm0.097$ & ---                                             & (12) \\
---                         & $2519\pm241$           & ---                &$692\pm42$        & $-3.152\pm0.272$        & $-4.677\pm0.313$ & ---                                             & (13) \\
0.461                       & 2368                   & ---                & 479              & $-3.154\pm0.084$       & $-4.675\pm0.097$ & ---                                             & (15) \\
---                         & $2344^{+718}_{-445}$   &  ---               & ---              & $-3.154\pm0.084$       & $-4.675\pm0.097$ & ---                                             & (16) \\
$0.633\pm0.033$             & $2069\pm142$           & $-0.031\pm0.123$   & $323\pm139$      & $-3.152\pm0.130$       & $-4.672\pm0.121$ & ---                                             & (17) \\
$0.630\pm0.100$               & $1585\pm234$           &  $-0.320\pm0.240$    & $631\pm260$      & $-3.170\pm0.110$         & $-4.690\pm0.130$   & ---                                             & (21) \\
0.608                       & 2369                   & ---                & 254              & $-3.145\pm0.075$       & $-4.690\pm0.097$  & $40.83\pm43.75$                                & (19) \\

  \hline
  \hline
 \end{tabular}
    \\
    \begin{minipage}{18cm}
     (1) \citet{Turner93}, (2) \citet{Loktin03}, (3) \citet{Kharchenko05}, (4) \citet{Wu09}, (5) \citet{Kharchenko09}, (6) \citet{VandePutte10}, (7) \citet{Kharchenko13}, (8) \citet{Dias14}, (9) \citet{Loktin17}, (10) \citet{Dias18}, (11) \citet{Soubiran18}, (12) \citet{Cantat-Gaudin18}, (13) \citet{Liu19}, (14) \citet{Bossini19}, (15) \citet{Cantat-Gaudin_Anders20}, (16) \citet{Cantat-Gaudin20}, (17) \citet{Dias21}, (18) \citet{Almeida23}, (19) \citet{Hunt23}, (20) \citet{Bonatto10}, (21) \citet{Angelo21}
     \end{minipage}
  \label{tab:literature}%
\end{table*}%

The intent of this study is to make a detailed, methodical analysis of both OCs with the aim of providing well-constrained estimates of the cluster parameters based on multiple independent methods. In this study, we analyzed the Roslund 3 and Ruprecht 174 OCs by determining the membership probabilities of their stars, mean proper motions, and distances. Our analysis combines ground-based {\it UBV} photometric observations with high-precision astrometry and photometry data from the {\it Gaia} third data release \citep[{\it Gaia} DR3,][]{Gaia23}. We report the fundamental parameters, luminosity and mass functions, the dynamical properties of mass segregation, and the kinematic and Galactic orbital properties of both OCs.

\begin{figure*}
\centering
\includegraphics[scale=0.8, angle=0]{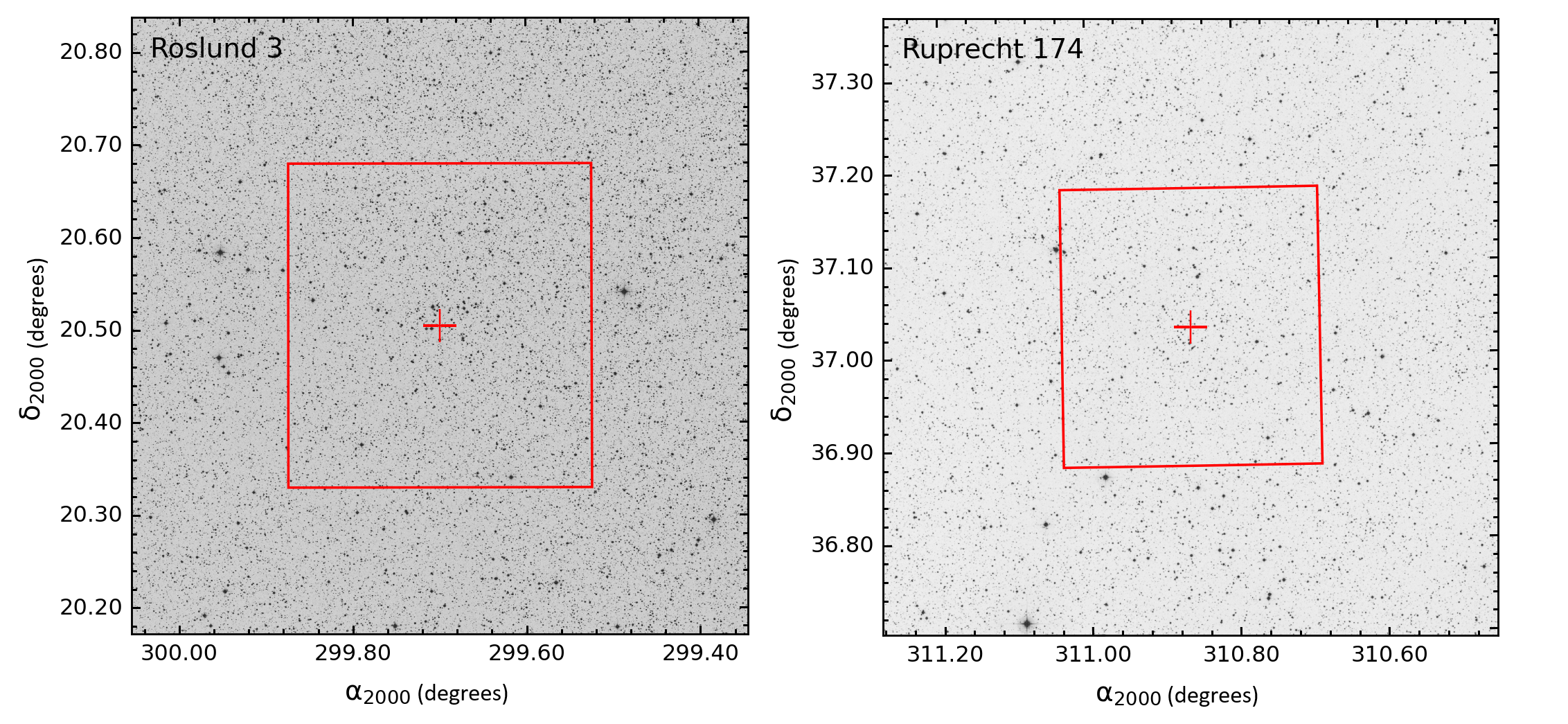}
\caption{Star fields for Roslund 3 (left panel) and Ruprecht 174 (right panel) in the equatorial coordinate system. The red boundaries indicate the fields observed with the T100 telescope, that is, $21'.5\times21'.5$. Each chart is $40'\times40'$.} North and east correspond to the up and left directions, respectively.
\label{fig:charts}
\end {figure*}


\begin{table}[ht]
\setlength{\tabcolsep}{10pt}
\renewcommand{\arraystretch}{1}
  \centering
  \caption{The observational data include information on both OCs. The table presents the cluster names, observation dates (day-month-year), and the filters employed in the {\it UBV} photometric system. For each filter, the exposure times (in seconds) and the corresponding number of exposures ($N$) are provided in separate rows.}
  \medskip
    \begin{tabular}{ccc}
\hline
\hline  
  \multicolumn{3}{c}{Filter/Exp, Time (s) $\times N$}   \\
    \hline
$U$           & $B$          & $V$          \\
    \hline
    \hline
  \multicolumn{3}{c}{Roslund 3}   \\
  \multicolumn{3}{c}{Observation Date: 23-07-2020}\\
   \hline
   \hline
   80$\times$1, 70$\times$3 &   8$\times$4 &   5$\times$4,10$\times$1 \\
700$\times$2 & 80$\times$4 & 60$\times$5 \\
1600$\times$3 & 900$\times$3 & 600$\times$2 \\
\hline
\hline
  \multicolumn{3}{c}{Ruprecht 174}   \\
  \multicolumn{3}{c}{Observation Date: 31-08-2022}\\
\hline
\hline
 15$\times$4 &  1.5$\times$4 &   6$\times$1 \\
 150$\times$6 & 15$\times$6 & 8$\times$6 \\
1500$\times$5 & 200$\times$5 & 150$\times$4 \\
    \hline
    \end{tabular}%
  \label{tab:exposures}%

\end{table}%


\section{Observations}
The CCD {\it UBV} photometric observations of Roslund 3 and Ruprecht 174 were collected at the TÜBİTAK National Observatory (TUG)\footnote{www.tug.tubitak.gov.tr} using the 1-m $f/10$ Ritchey-Chrétien telescope (T100). Observations were performed on 2020 July 23 for Roslund 3 and on 2022 August 31 for Ruprecht 174. The T100 was equipped with a back-illuminated 4k$\times$4k pixel Fairchild CCD camera, optimized for enhanced UV sensitivity. This setup provided an image scale of $0''\!.31$ pixel$^{-1}$ and a total field of view of $21'\!.5 \times 21'\!.5$.

To ensure optimal signal-to-noise ratios for faint stars while preventing saturation of bright stars, multiple exposure times were employed in each filter. Identification charts for the OCs are shown in Figure~\ref{fig:charts}, and a detailed observation log is provided in Table~\ref{tab:exposures}. For Roslund 3, a total of 32 images were obtained with exposure times ranging from 5s to 1600s in {\it U, B,} and {\it V} bands. Similarly, 41 images were collected for Ruprecht 174 with exposure times between 1.5s and 1500s. Photometric calibration was performed using standard stars from \citet{Landolt09}, observed at various airmasses alongside the cluster fields. Across two observing nights, 99 standard stars from 18 Landolt fields were recorded, covering an airmass range of $1.231 \leq X \leq 1.979$ on 2020 July 23 and $1.012 \leq X \leq 2.455$ on 2022 August 31 (Table~\ref{tab:standard_stars}). Technical details regarding the observing conditions at the TUG observing site are provided in \citet{Ak24}. Standard CCD calibrations, including bias and flat-field corrections, as well as the determination of instrumental magnitudes for standard stars \citep{Landolt09}, were performed using the relevant IRAF\footnote{IRAF was distributed by the National Optical Astronomy Observatories} packages. Specifically, the instrumental magnitudes of the standard stars were measured applying the aperture photometry routines within IRAF. Multiple linear regression analysis was performed on instrumental magnitudes of standard stars to derive photometric extinction and transformation coefficients for each observing night, presented in Table~\ref{tab:coefficients}. The instrumental magnitudes of stars in the cluster fields were measured using the Point Spread Function (PSF) method \citep{Stetson87}. After this, the instrumental magnitudes were transformed into standard {\it UBV} magnitudes and colors using the transformation equations provided by \citet{Janes11}:

\begin{align}\label{eq:01}
v =& V + \alpha_{\rm bv}(B-V) + k_{\rm v}X + C_{\rm bv}, \\ \nonumber
b =& V + \alpha_{\rm b}(B-V) + k_{\rm b}X + k'_{\rm b}X(B-V) + C_{\rm b}, \\ \nonumber
u =& V + (B-V) + \alpha_{\rm ub}(U-B) + k_{\rm u}X + k'_{\rm u}X(U-B) \\ \nonumber
&+ C_{\rm ub}
\end{align}
In the Equation (\ref{eq:01}), $U$, $B$ and $V$ represent the magnitudes in the standard photometric system, while $u$, $b$ and $v$ denote the instrumental magnitudes. The parameter $X$ corresponds to the airmass, whereas $k$ and $k'$ represent the primary and secondary extinction coefficients, respectively. $C_{\rm i}$ and $\alpha$ are usually called zeropoint and color term coefficient, respectively.

\begin{table}[h]
\setlength{\tabcolsep}{5pt}
\renewcommand{\arraystretch}{0.8}
  \centering
\caption{Selected standard star fields from \citet{Landolt09}. The table columns indicate the observation date (day-month-year), the name of the standard star field from Landolt, the number of standard stars ($N_{\rm st}$) observed in each field, the number of observations per field ($N_{\rm obs}$), and the airmass range ($X$) during which the observations were conducted.}
\medskip
    \begin{tabular}{lcc}
    \hline 
    \multicolumn{3}{c}{Observation Date: 23.07.2020}\\
     \multicolumn{3}{c}{Air Mass: 1.231 -- 1.979}\\
    \hline
    \hline
Star Field	& $N_{\rm st}$ & $N_{\rm obs}$\\
\hline
SA93      &  4 	  & 1	\\
SA106     &  2	  & 1	\\
SA107	  &  7	  & 1	\\
SA108     &  2	  & 1	\\
SA110SF2  & 10	  & 1   \\
SA111	  &  5	  & 2	\\
SA112	  &  6	  & 1	\\
SA113	  & 15	  & 2	\\
SA114	  &  5	  & 1	\\
\hline	 
\multicolumn{3}{c}{Observation Date: 31.08.2022}\\
     \multicolumn{3}{c}{Air Mass: 1.012 -- 2.455}\\
     \hline
\hline
SA20SF3   &  7    & 2	\\
SA23SF1   &  9    & 1	\\
SA38SF1   &  8    & 1	\\
SA41      &  5	  & 1	\\
SA92SF3	  &  6 	  & 2	\\
SA94      &  2    & 1	\\
SA96	  &  2	  & 1   \\
SA106     &  2	  & 1	\\
SA108SF1  &  2	  & 1	\\
SA111	  &  5	  & 1	\\
SA112	  &  6	  & 2	\\	
SA113	  &  15	  & 1	\\
SA114	  &  5	  & 2	\\	       
    \hline \end{tabular}%
  \label{tab:standard_stars}%
\end{table}%

\begin{table*}
\renewcommand{\arraystretch}{1}
  \centering
\caption{Extinction and transformation coefficients derived for the two observation nights: $k$ and $k'$ represent the first- and second-order extinction coefficients, respectively, while $\alpha$ and $C_{\rm i}$ correspond to the transformation coefficients. Dates are given in the format day-month-year.} 
  \medskip
    \begin{tabular}{lcccc}
    \hline
    & \multicolumn{4}{c}{Observation Date: 23.07.2020}\\
    \hline
Filter/color & $k$            & $k'$                  & $\alpha$              & $C_{\rm i}$                 \\
    \hline
$U$     & $0.493 \pm 0.116$ & $-0.272 \pm 0.107$  & ---                 & ---               \\
$B$     & $0.272 \pm 0.058$ & $-0.097 \pm 0.050$  & $1.044 \pm 0.081$   & $2.130 \pm 0.088$ \\
$V$     & $0.135 \pm 0.022$ & ---                 & ---                 & ---               \\
$U-B$   &  ---              & ---                 & $1.221 \pm 0.161$   & $4.491 \pm 0.169$ \\
$B-V$   &  ---              & ---                 & $0.066 \pm 0.010$   & $2.174 \pm 0.033$ \\
\hline
    & \multicolumn{4}{c}{Observation Date: 31.08.2022}\\
\hline
$U$     &  $0.470 \pm 0.067$ & $-0.005 \pm 0.056$ & ---                 & ---                \\ 
$B$     &  $0.263 \pm 0.045$ & $-0.025 \pm 0.044$ & $0.956 \pm 0.059$   & $0.558 \pm 0.058$  \\
$V$     &  $0.182 \pm 0.014$ & ---                & ---                 & ---                \\
$U-B$   &  ---               & ---                & $0.842 \pm 0.074$   & $2.916 \pm 0.084$  \\
$B-V$   &  ---               & ---                & $0.065 \pm 0.008$   & $0.053 \pm 0.0190$ \\
\hline
    \end{tabular}%
  \label{tab:coefficients}%
\end{table*}%

\section{Data Analysis}
\subsection{Photometric data}
{\it UBV}-based photometric catalogs were constructed for stars within the regions of Roslund 3 and Ruprecht 174, including their equatorial coordinates ($\alpha, \delta$), $V$ magnitudes, and $U-B$ and $B-V$ color indices, along with their uncertainties. The catalog for Roslund 3 comprises data for 3,957 stars with $V$-band apparent magnitudes in the range $10 < V~{\rm (mag)}< 19.5$, whereas the catalog for Ruprecht 174 includes 6,999 stars with magnitudes within $9 < V~{\rm (mag)}< 21.5$. In addition to the {\it UBV} catalogs obtained from observations, {\it Gaia} DR3 data \citep{Gaia23} were separately compiled for each cluster to utilize a comprehensive analysis. These catalogs are utilized to determine the membership probabilities of stars and to derive astrometric parameters as well as astrophysical parameters based on recent {\it Gaia} data. The {\it Gaia} catalogs for the two OCs encompass stars within $40' \times 40'$ regions centered on each cluster.

These datasets contain photometric measurements ($G$, $G_{\rm BP}$, $G_{\rm RP}$), proper-motion (PM) components ($\mu_{\alpha}\cos\delta$, $\mu_\delta$), trigonometric parallaxes ($\varpi$), radial velocities ($V_{\rm R}$), and the associated uncertainties. These are listed in Table~\ref{tab:all_cat}. The number of stars detected in the {\it Gaia} catalog for the Roslund 3 and Ruprecht 174 is 377,204 and 193,424, respectively. These stars are in the magnitude range $6< G~{\rm (mag)}<22.5$ for both clusters. Figure~\ref{fig:charts} illustrates the $40' \times 40'$ cluster fields as observed in the {\it Gaia} system, with the coverage of {\it UBV} data outlined in red. 

To accurately determine the astrophysical parameters of the clusters, photometric measurements were first analyzed to establish the faint magnitude limits, known as photometric completeness limits, of stars within the cluster regions. These values indicate the completeness level of the data, corresponding to the faintest magnitudes at which stars are still fully recovered in the analysis. They essentially mark the point beyond which the detection rate of objects drops significantly.  The same approach was applied to determine the photometric completeness limits in both the {\it UBV} and {\it Gaia} photometric systems.

\begin{figure}
\centering
\includegraphics[width=\columnwidth]{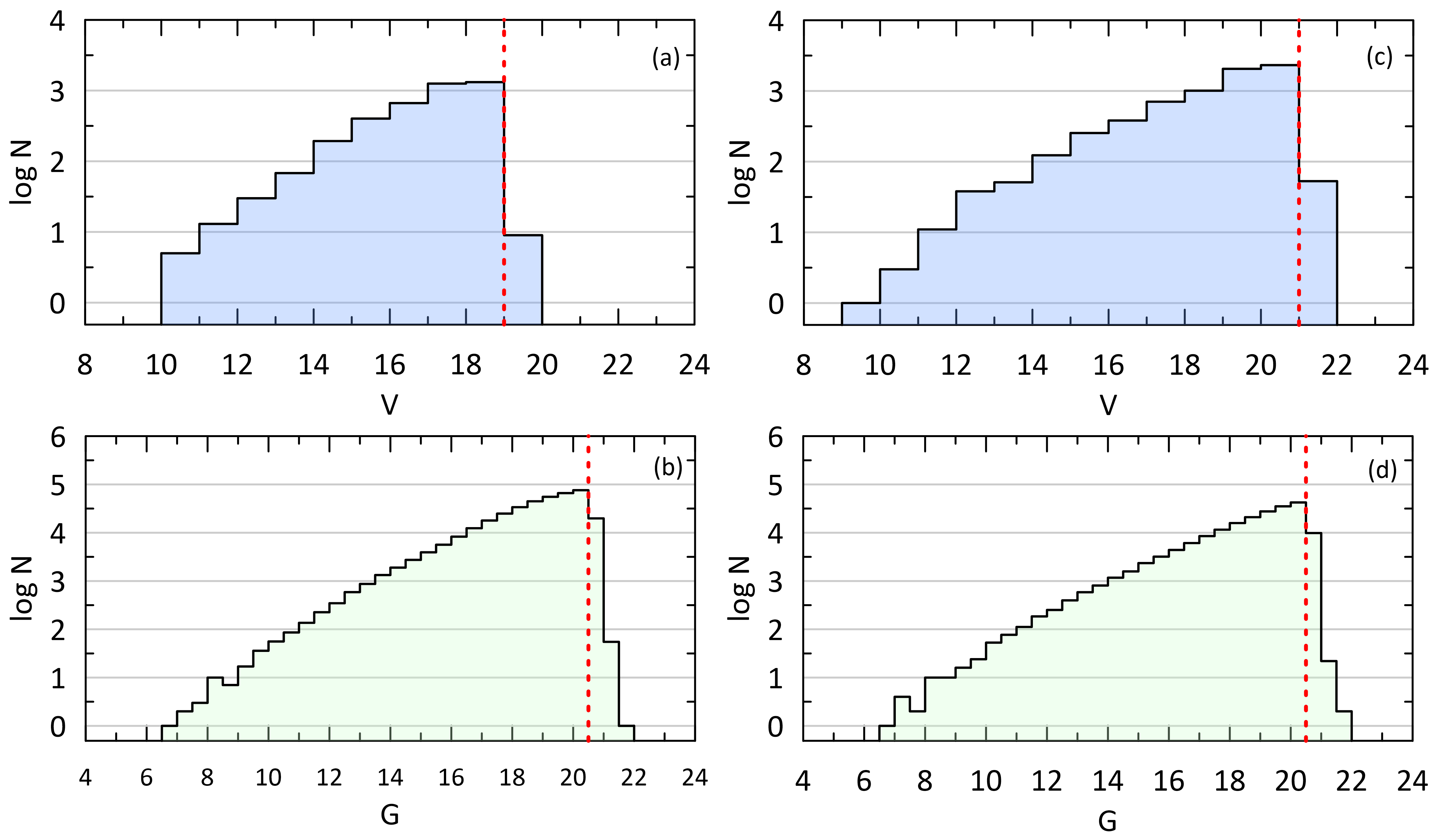}\\
\caption{The histograms illustrate the distribution of stars in the direction of Roslund 3 (panels a and b) and Ruprecht 174 (panels c and d) as a function of $V$ and $G$ magnitudes. The red dashed lines indicate the photometric completeness limits for each band.} 
\label{fig:histograms}
\end {figure}

Histograms picturing star counts as a function of $V$ and $G$-apparent magnitudes were generated for the stars included in both catalogs. The completeness limits for $V$ and $G$ magnitudes were defined as the magnitude at which the star count reaches its peak in these distributions. Figure~\ref{fig:histograms} presents these histograms for the cluster fields. Panels (a) and (c) display the star count distributions as a function of $V$ magnitude for Roslund 3 and Ruprecht 174, respectively. In the case of Roslund 3 (Figure~\ref{fig:histograms}a), the number of stars increases toward fainter magnitudes, peaking at $V$=19 mag, beyond which a sharp decline is observed. Whereas for Ruprecht 174, star counts peak at $V$=21 mag. Panels (b) and (d) illustrate the distribution of stars in terms of $G$ magnitude, binned at 0.5 mag intervals, for Roslund 3 and Ruprecht 174. The maximum star count occurs at $G$=20.5 mag for both clusters.

The uncertainties in apparent magnitudes and color indices derived from the {\it UBV} and {\it Gaia} photometric systems are listed in Table~\ref{tab:photometric_errors} as a function of $V$ and $G$ magnitudes. Measurement errors primarily arise from statistical uncertainties, while errors in color indices result from error propagation within the respective photometric bands. For Roslund 3, considering the completeness limit of $V=19$ mag, the mean photometric errors do not exceed 0.020 mag, while the uncertainties in the $U-B$ and $B-V$ color indices reach up to 0.080 and 0.028 mag, respectively. In Ruprecht 174, with a completeness limit of $V=21$ mag, the corresponding mean errors are 0.045 mag for $V$, 0.105 mag for $U-B$, and 0.092 mag for $B-V$. An analysis of mean errors in the {\it Gaia} photometric system reveals uncertainties of 0.011 and 0.001 mag in $G$-band and 0.199 and 0.197 mag in $G_{\rm BP}-G_{\rm RP}$ for Roslund 3 and Ruprecht 174, respectively.

\begin{table*}[]
\caption{The catalogs for Roslund 3 and Ruprecht 174. The complete table can be found electronically.}
\resizebox{\textwidth}{!}{%
\begin{tabular}{cccccccccccc}
\hline
\multicolumn{11}{c}{Roslund 3}                                                                                                                                                              &       \\ \hline
ID   & RA            & DEC           & $V$           & $U-B$        & $B-V$        & $G$           & $G_{\rm BP}-G_{\rm RP}$ & $\mu_{\alpha}\cos\delta$ & $\mu_{\delta}$    & $\varpi$      & $P$   \\
     & (hh:mm:ss.ss) & (dd:mm:ss.ss) & (mag)         & (mag)        & (mag)        & (mag)         & (mag)                   & (mas yr$^{-1}$)          & (mas yr$^{-1}$)   & (mas)         &       \\ \hline
0001 & 19:58:03.17   & +20:23:16.81  & 18.742(0.034) & ...          & 1.229(0.047) & 19.673(0.005) & 1.566(0.076)            & $-03.362$(0.397)         & $-005.876$(0.357) & 0.611(0.379)  & 0.236 \\
0002 & 19:58:03.28   & +20:30:09.58  & 17.904(0.014) & ...          & 1.413(0.022) & 17.605(0.003) & 1.676(0.014)            & $-03.685$(0.091)         & $-005.498$(0.082) & 0.090(0.096)  & 0.000 \\
0003 & 19:58:03.29   & +20:32:17.12  & 17.011(0.007) & 0.607(0.039) & 1.138(0.009) & 16.857(0.003) & 1.352(0.008)            & $-07.901$(0.056)         & $-007.796$(0.052) & 0.356(0.060)  & 0.000 \\
0004 & 19:58:03.33   & +20:33:14.89  & 17.523(0.010) & 0.572(0.056) & 0.119(0.014) & 17.460(0.003) & 1.277(0.011)            & $-03.049$(0.078)         & $-004.778$(0.069) & 0.325(0.086)  & 0.000 \\
...  & ...           & ...           & ...           & ...          & ...          & ...           & ...                     & ...                      & ...               & ...           & ...   \\
...  & ...           & ...           & ...           & ...          & ...          & ...           & ...                     & ...                      & ...               & ...           & ...   \\
3954 & 19:59:30.36   & +20:40:06.53  & 17.294(0.009) & 0.858(0.060) & 1.124(0.012) & 17.158(0.003) & 1.347(0.009)            & $-05.195$(0.061)         & $-011.663$(0.058) & 0.508(0.062)  & 0.000 \\
3955 & 19:59:30.40   & +20:36:04.09  & 17.823(0.013) & 0.556(0.067) & 1.103(0.017) & 17.685(0.003) & 1.344(0.012)            & $-04.323$(0.087)         & $-010.618$(0.084) & 0.521(0.092)  & 0.089 \\
3956 & 19:59:30.41   & +20:36:29.56  & 16.293(0.024) & ---          & 1.771(0.065) & 15.583(0.003) & 2.162(0.006)            & $-02.562$(0.029)         & $-005.303$(0.028) & 0.044(0.032)  & 0.000 \\
3957 & 19:58:30.55   & +20:39:41.56  & 14.062(0.006) & 0.271(0.013) & 0.725(0.010) & 14.054(0.003) & 0.945(0.005)            & 05.029(0.015)            & $-006.371$(0.014) & 1.284(0.015)  & 0.000 \\ \hline
\multicolumn{11}{c}{Ruprecht 174}                                                                                                                                                           &       \\ \hline
ID   & RA            & DEC           & $V$           & $U-B$        & $B-V$        & $G$           & $G_{\rm BP}-G_{\rm RP}$ & $\mu_{\alpha}\cos\delta$ & $\mu_{\delta}$    & $\varpi$      & $P$   \\
     & (hh:mm:ss.ss) & (dd:mm:ss.ss) & (mag)         & (mag)        & (mag)        & (mag)         & (mag)                   & (mas yr$^{-1}$)          & (mas yr$^{-1}$)   & (mas)         &       \\ \hline
0001 & 20:42:28.16   & +37:09:50.19  & 19.649(0.074) & ...          & ...          & 19.719(0.005) & 1.771(0.074)            & $-01.685$(0.306)         & $-003.678$(0.408) & -0.696(0.344) & 0.000 \\
0002 & 20:42:28.25   & +37:09:36.01  & 19.997(0.044) & ...          & 1.470(0.083) & 19.407(0.004) & 1.688(0.059)            & $-02.024$(0.230)         & $-003.361$(0.317) & 0.144(0.255)  & 0.002 \\
0003 & 20:42:28.81   & +37:09:16.45  & 19.903(0.037) & 0.574(0.087) & 1.242(0.059) & 19.319(0.004) & 1.721(0.053)            & $-02.453$(0.221)         & $-003.977$(0.307) & 0.042(0.244)  & 0.026 \\
0004 & 20:42:28.92   & +37:09:00.94  & 19.337(0.020) & 0.509(0.052) & 1.343(0.035) & 18.797(0.003) & 1.657(0.030)            & $-01.419$(0.150)         & $-003.508$(0.225) & 0.196(0.171)  & 0.000 \\
...  & ...           & ...           & ...           & ...          & ...          & ...           & ...                     & ...                      & ...               & ...           & ...   \\
...  & ...           & ...           & ...           & ...          & ...          & ...           & ...                     & ...                      & ...               & ...           & ...   \\
6996 & 20:44:17.36   & +36:54:24.66  & 19.253(0.017) & 1.010(0.065) & 1.500(0.032) & 18.613(0.004) & 1.973(0.037)            & $-02.194$(0.162)         & $-006.348$(0.195) & 0.365(0.191)  & 0.015 \\
6997 & 20:44:17.44   & +36:52:13.10  & 14.693(0.009) & 0.246(0.017) & 0.699(0.014) & 14.496(0.003) & 0.890(0.005)            & $-01.048$(0.015)         & $-008.408$(0.019) & 1.185(0.018)  & 0.000 \\
6998 & 20:44:17.62   & +36:53:23.61  & 17.021(0.011) & 0.584(0.007) & 1.122(0.011) & 16.682(0.003) & 1.371(0.007)            & $-02.003$(0.043)         & $-006.644$(0.051) & 0.712(0.052)  & 0.000 \\
6999 & 20:44:18.11   & +36:52:45.79  & 19.390(0.020) & 0.645(0.060) & 1.484(0.036) & 19.687(0.004) & 1.854(0.078)            & $-02.876$(0.290)         & $-006.148$(0.354) & -0.189(0.330) & 0.000 \\ \hline
\end{tabular}%
}\label{tab:all_cat}
\end{table*}

\begin{table*}
  \centering
\setlength{\tabcolsep}{4pt}
\renewcommand{\arraystretch}{0.8}
\caption{Mean internal photometric uncertainties corresponding to each magnitude bin in the $V$ and $G$ bands.}
    \begin{tabular}{ccccccccc}
      \hline
    \multicolumn{5}{c}{Roslund 3} & \multicolumn{4}{c}{Ruprecht 174} \\
    \hline
  $V$ & $N$ & $\sigma_{\rm V}$ & $\sigma_{\rm U-B}$ & $\sigma_{\rm B-V}$ & $N$ & $\sigma_{\rm V}$ & $\sigma_{\rm U-B}$ & $\sigma_{\rm B-V}$\\
  \hline
[09, 12] &   18 & 0.002 & 0.003 & 0.003 &   15 & 0.002 & 0.004 & 0.003 \\
(12, 14] &   98 & 0.004 & 0.015 & 0.007 &   89 & 0.004 & 0.010 & 0.007 \\
(14, 15] &  193 & 0.008 & 0.032 & 0.014 &  125 & 0.009 & 0.022 & 0.014 \\
(15, 16] &  402 & 0.015 & 0.068 & 0.027 &  252 & 0.014 & 0.027 & 0.021 \\
(16, 17] &  666 & 0.014 & 0.075 & 0.026 &  382 & 0.008 & 0.013 & 0.010 \\
(17, 18] & 1256 & 0.010 & 0.040 & 0.014 &  706 & 0.015 & 0.026 & 0.020 \\
(18, 19] & 1316 & 0.020 & 0.080 & 0.028 & 1008 & 0.009 & 0.032 & 0.017 \\
(19, 20] &    8 & 0.038 & ---   & 0.053 & 2055 & 0.021 & 0.064 & 0.040 \\
(20, 22] & ---  & ---   & ---   & ---   & 2367 & 0.045 & 0.105 & 0.092 \\
  \hline
  $G$ & $N$ & $\sigma_{\rm G}$ &   \multicolumn{2}{c}{$\sigma_{G_{\rm BP}-G_{\rm RP}}$} &$N$ & $\sigma_{\rm G}$ & \multicolumn{2}{c}{$\sigma_{G_{\rm BP}-G_{\rm RP}}$}\\
  \hline
  [05, 10] &     40 & 0.003 & \multicolumn{2}{c}{0.008} & 43    & 0.003 & \multicolumn{2}{c}{0.007} \\
  (10, 12] &    314 & 0.003 & \multicolumn{2}{c}{0.006} & 266   & 0.003 & \multicolumn{2}{c}{0.007} \\
  (12, 14] &   2031 & 0.003 & \multicolumn{2}{c}{0.006} & 1420  & 0.003 & \multicolumn{2}{c}{0.006} \\
  (14, 15] &   3223 & 0.003 & \multicolumn{2}{c}{0.006} & 1979  & 0.003 & \multicolumn{2}{c}{0.006} \\
  (15, 16] &   6660 & 0.003 & \multicolumn{2}{c}{0.006} & 3927  & 0.003 & \multicolumn{2}{c}{0.006} \\
  (16, 17] &  13925 & 0.003 & \multicolumn{2}{c}{0.009} & 7596  & 0.003 & \multicolumn{2}{c}{0.009} \\
  (17, 18] &  30270 & 0.003 & \multicolumn{2}{c}{0.016} & 14627 & 0.003 & \multicolumn{2}{c}{0.016} \\
  (18, 19] &  58730 & 0.004 & \multicolumn{2}{c}{0.032} & 27407 & 0.003 & \multicolumn{2}{c}{0.036} \\
  (19, 20] & 100213 & 0.005 & \multicolumn{2}{c}{0.073} & 48697 & 0.004 & \multicolumn{2}{c}{0.073} \\
  (20, 21] & 141914 & 0.011 & \multicolumn{2}{c}{0.199} & 77592 & 0.001 & \multicolumn{2}{c}{0.197} \\
  (21, 23] &  19884 & 0.028 & \multicolumn{2}{c}{0.435} & 9870  & 0.026 & \multicolumn{2}{c}{0.423} \\
      \hline
    \end{tabular}%
  \label{tab:photometric_errors}%
\end{table*}%


\subsection{The Radial Density Profile}
The analysis of radial density profiles (RDPs) aims to investigate the spatial distribution and structural properties of OCs. The RDP of a cluster is defined by the variation in stellar number density as a function of the radial distance from the cluster center. Typically, the highest stellar density is observed in the central region of the cluster, gradually decreasing outward until it merges with the background stellar population. The analysis of these variations facilitates the determination of fundamental physical parameters, including the total stellar content, the mass distribution, and the luminosity of the cluster. 

\begin{figure}[htbp]
\centering
\includegraphics[scale=0.25, angle=0]{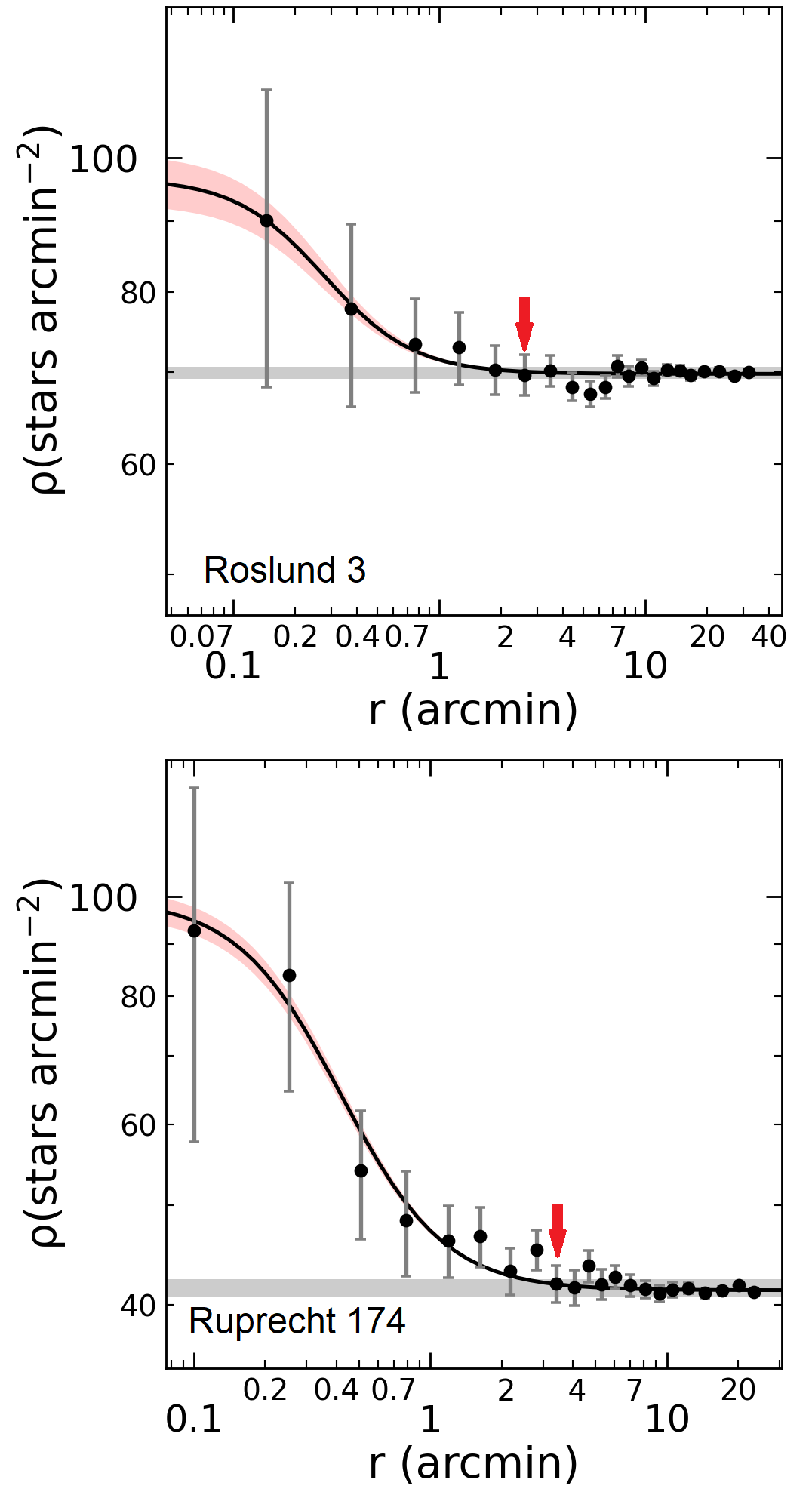}\\
\caption{King profiles of Roslund 3 (upper panel) and Ruprecht 174 (lower panel), with black solid curves representing the best-fit models. The shaded red and gray regions indicate the confidence intervals of the model and the background stellar density, respectively, while the red arrows show the limiting radii of the OCs.}
\label{fig:king}
\end {figure}

The structural parameters of the RDP analysis were derived using photometric $G$ magnitudes obtained from the {\it Gaia} DR3 dataset. The central coordinates of the clusters, referenced in the equatorial coordinate system, were adopted from the catalog of \citet{Cantat-Gaudin20}. To construct the RDPs, stars were counted within concentric annuli of varying radii, measured from the cluster center. The stellar number density for each annulus was subsequently calculated using the formula $\rho(r_i) = N_i / A_i$, where $N$ represents the count of stars within a given annulus, and $A$ corresponds to its area. The subscript $i$ indicates the specific annulus under consideration. The observed RDPs were then compared with the empirical model proposed by \citet{King62}, which describes the density distribution of star clusters. The model is expressed as follows: $\rho(r)=f_{\rm bg}+ [f_{\rm 0}/(1+(r/r_{\rm c})^2)]$; where $r$ is the radius from the cluster center, $f_{\rm 0}$ is the central stellar density, $r_{\rm c}$ is the core radius, and $f_{\rm bg}$ is the background stellar density. A chi-squared minimization method was applied to determine the best-fitting model parameters. Final parameter estimates were obtained by selecting the values that yielded the minimum $\chi^2$. It has been visually determined that the stellar density profiles of Roslund 3 and Ruprecht 174 are well represented by the King model. The resulting RDPs for the analyzed clusters, together with the best-fit models from \citet{King62}, are presented in Figure \ref{fig:king}. The derived structural parameters are summarized in Table~\ref{tab:Final_table}. The point at which the observational and model density profiles converge, approximately 2.5 arcminutes for Roslund~3 and 3.5 arcminutes for Ruprecht~174 from their respective centers, has been adopted as the limiting radius ($r_{\rm lim}^{\rm obs}$) for each OC. These distances represent the radii beyond which the cluster population blends into the field star background.

The concentration parameter, defined as $C = \log(r_{\rm lim}/r_{\rm c})$, is a key indicator of the internal structure and stellar density distribution of OCs \citep[e.g.][]{King62, Tadross21, Badawy24, Alzahrani25}. It aids in differentiating clusters from the field population by quantifying their degree of central condensation \citep{Richstone86}. In this study, $C$ was found to be $0.990\pm 0.093$ for Roslund 3 and $1.018\pm 0.014$ for Ruprecht 174, indicating similarly moderately concentrated structures. Results suggest that both clusters have undergone partial dynamical evolution with some degree of core development \citep{Elson87}.

\subsection{Membership Selection}
\label{section:membership}
We estimated the membership probabilities of stars located within the fields of the investigated OCs using the {\sc UPMASK} algorithm \citep{Krone-Martins14}. For Roslund 3 and Ruprecht 174, astrometric and positional data from the {\it Gaia} DR3 catalog were used to construct a five-dimensional input space comprising equatorial coordinates ($\alpha$, $\delta$), trigonometric parallax ($\varpi$), PM components ($\mu_{\alpha} \cos \delta$, $\mu_{\delta}$), and the corresponding uncertainties for each star (see Table~\ref{tab:all_cat}). The {\sc UPMASK} procedure was implemented with varying $k$-means cluster numbers, ranging from 6 to 30, and each configuration was iterated 25 times. The optimal number of clusters was selected based on the configuration that most effectively captured the spatial and kinematic structure of each cluster through membership probability distributions. As a result, $k$ values of 8 and 7 were adopted for Roslund 3 and Ruprecht 174, respectively. To ensure well-defined sequences in the color-magnitude diagrams (CMDs) and to improve the reliability of the derived astrophysical and kinematic parameters, only stars with membership probabilities $P \geq 0.5$ were retained.

The membership probabilities of stars derived from {\it Gaia} DR3 astrometric data were also adopted as the membership probabilities for the corresponding stars identified in the {\it UBV} catalogs. By matching the equatorial coordinates of stars in both the {\it Gaia} and {\it UBV} datasets, the probabilities calculated from {\it Gaia} parameters were assigned to their counterparts in the {\it UBV} photometric catalog (see Table~\ref{tab:all_cat}). This approach ensured a consistent membership classification across both datasets.

While statistical methods are commonly applied to estimate membership probabilities, their reliability can be affected by factors such as line-of-sight reddening, increased photometric errors at fainter magnitudes, and the presence of unresolved binaries along the main-sequence. To address these challenges, cluster membership was determined by considering both astrometric and photometric approaches. The positions of stars with $P \geq 0.5$ on the $G$ vs $(G_{\rm BP}-G_{\rm RP})$ CMDs were analyzed to visually identify the lower boundary of the cluster main sequence. This boundary was defined by shifting solar metallicity ($Z$=0.0152) zero-age main sequence (ZAMS) of {\sc PARSEC} models  \citep{Bressan12} through the $G$ magnitude and $G_{\rm BP}-G_{\rm RP}$ color to account for unresolved binaries, reddening, and observational scatter. This method enabled the identification of probable main-sequence members and bright stars with high membership probabilities near the turn-off and giant regions. For Roslund 3 and Ruprecht 174, the number of stars with $P \geq 0.5$ and those limited with {\sc PARSEC} isochrone boundaries was found to be 198 and 397, respectively. These selected stars were subsequently used in {\it Gaia}-based analyses for deriving astrometric and astrophysical parameters.

In the {\it UBV} system, the lower main-sequence boundary was determined using the observational ZAMS curve of \citet{Sung13} with solar metallicity, which was empirically shifted in both $V$ magnitude and $B-V$ color with the same approaches as used in \textit{Gaia}-based CMD. For Roslund 3 and Ruprecht 174, the number of stars with $P \geq 0.5$ within the observational ZAMS boundaries from the {\it UBV} system was found to be 78 and 106, respectively. These stars were subsequently utilized in the determination of cluster parameters such as reddening, metallicity, distance, and age within the {\it UBV} framework. By standardizing the selection process across both photometric systems, this approach facilitated the identification of the most likely main-sequence members and bright stars with high membership probabilities near the cluster’s turn-off point and giant regions, enhancing the consistency and reliability of the cluster member selection.

\begin{figure}[t]
\centering
\includegraphics[width=0.95\columnwidth]{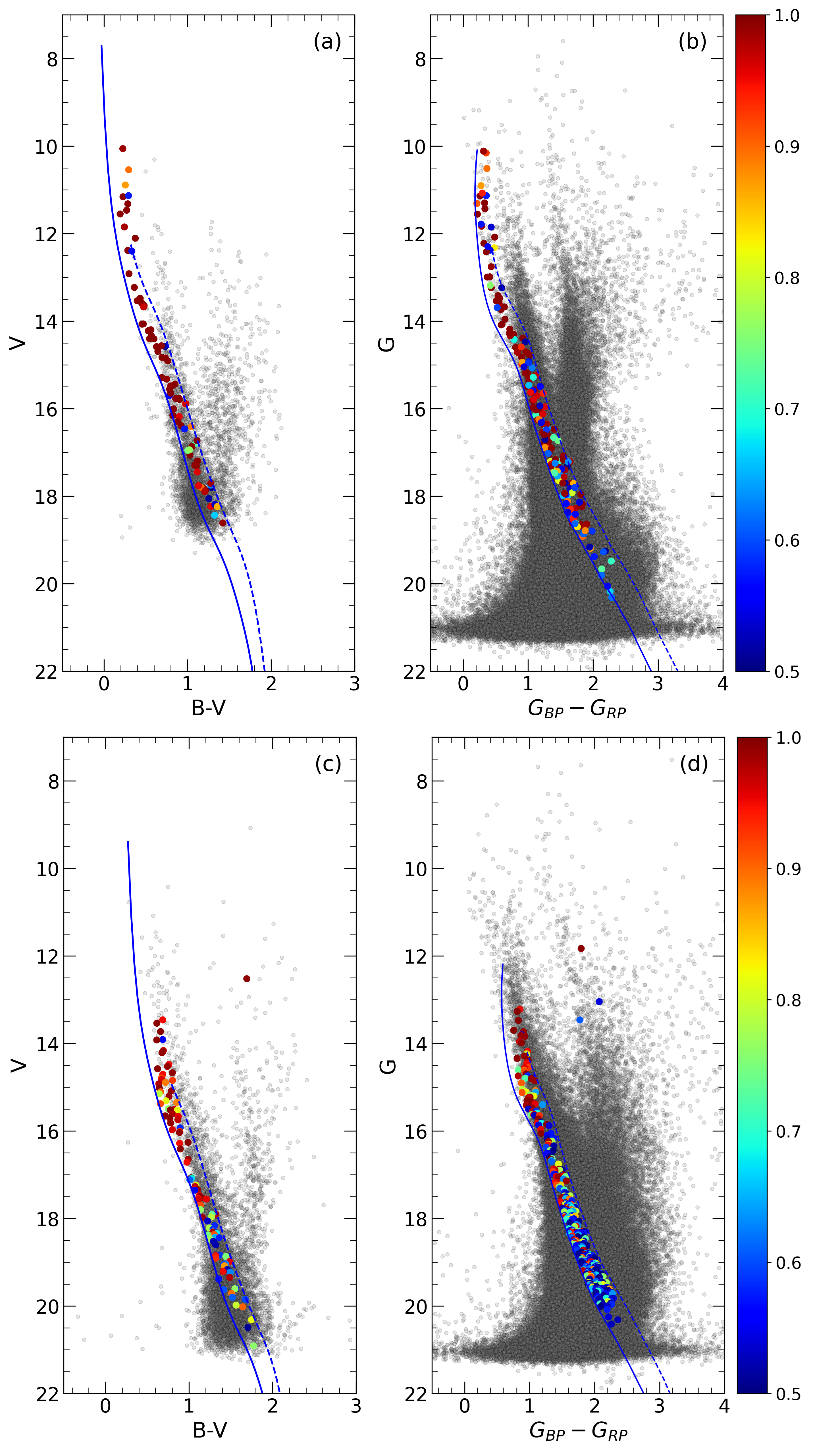}
\caption{CMDs of the Roslund 3 (a-b) and Ruprecht 174 (c-d) OCs are shown in both the {\it UBV} (a and c) and {\it Gaia} (b and d) photometric systems. Stars with membership probabilities $P < 0.5$ are represented by gray circles, while those with $P \geq 0.5$ are denoted by colored circles. The continuous and dashed blue curves correspond to the ZAMS lines from \citet{Sung13} and PARSEC models \citep{Bressan12}.}
\label{fig:CMD}
\end {figure}

Figure~\ref{fig:CMD} displays the distribution of stars in both the {\it UBV} and {\it Gaia}-based CMDs. The ZAMS of \citet{Sung13} and PARSEC restricted bands are overplotted as solid and dashed blue curves, respectively. The $B$ vs $(B-V)$ CMDs for stars identified from {\it UBV} data are presented in Figures~\ref{fig:CMD}a and \ref{fig:CMD}c for Roslund 3 and Ruprecht 174, respectively. Similarly, the $G$ vs $(G_{\rm BP} - G_{\rm RP})$ CMDs are shown in Figures~\ref{fig:CMD}b and \ref{fig:CMD}d for Roslund 3 and Ruprecht 174, respectively. In both \textit{UBV} and \textit{Gaia}-based CMDs, stars are color-coded based on their membership probabilities, with those having $P\geq 0.5$ highlighted.

\begin{figure}[!t]
\centering
\includegraphics[width=\columnwidth]{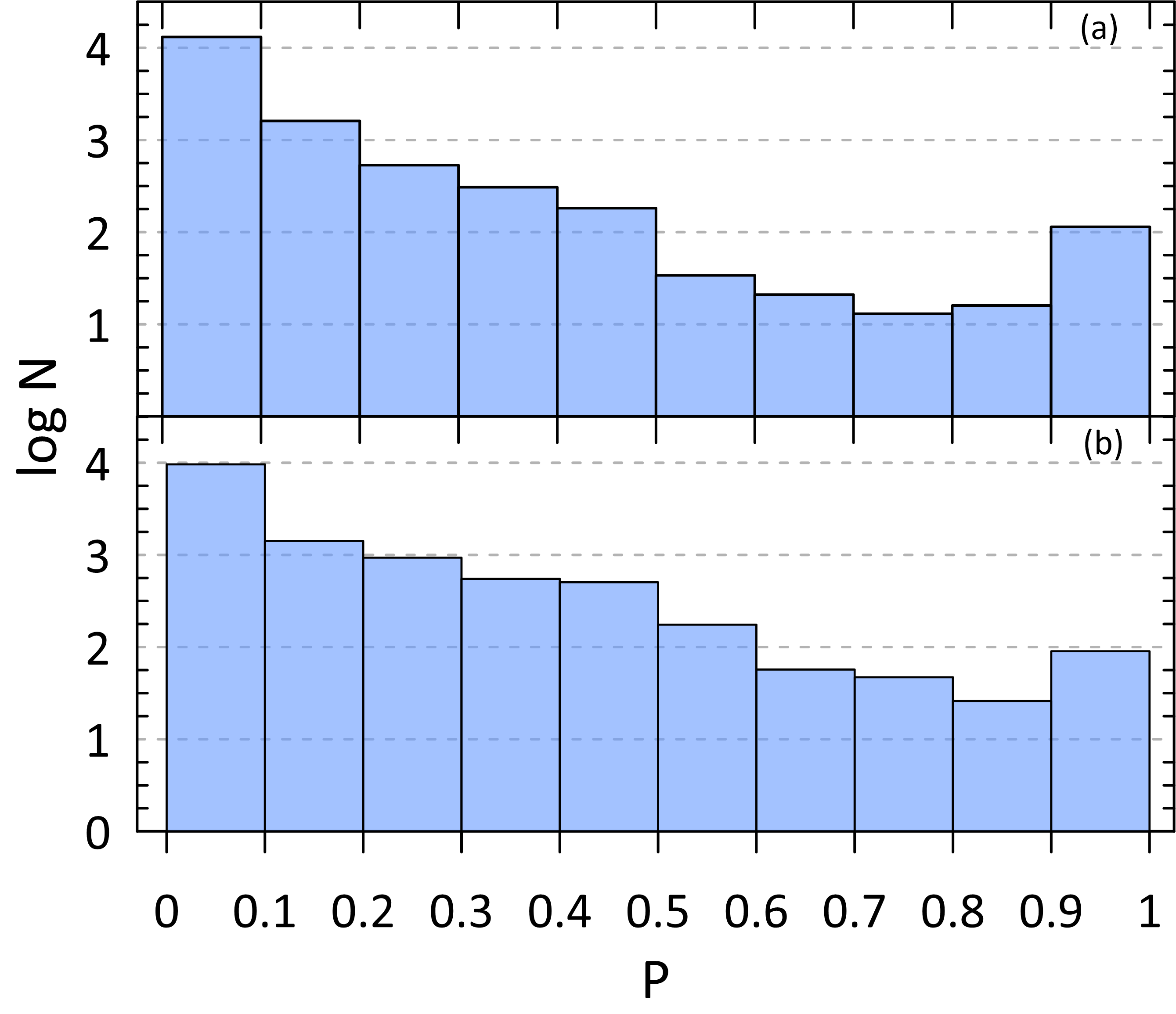}
\caption{The membership probability distributions for all stars in the cluster fields of Roslund 3 (a) and Ruprecht 174 (b) are shown, based on {\it Gaia} data. The blue histograms represent the membership probabilities of all stars detected in the cluster regions.}
\label{fig:prob_hists} 
\end {figure}

Figure~\ref{fig:prob_hists} presents the probability histograms of stars identified in the {\it Gaia}-based data, including only those brighter than the $G$ photometric completeness limits. An analysis of stars with $P \geq 0.5$ within the defined boundary radii revealed that their proportions relative to the total stellar population in the $40' \times 40'$ cluster area ($0 \leq P \leq 1$) were $1.2\%$ for Roslund 3 and $2.9\%$ for Ruprecht 174, suggesting that neither cluster is significantly crowded.

\begin{figure*}[ht]
\centering
\includegraphics[width=1.6\columnwidth]{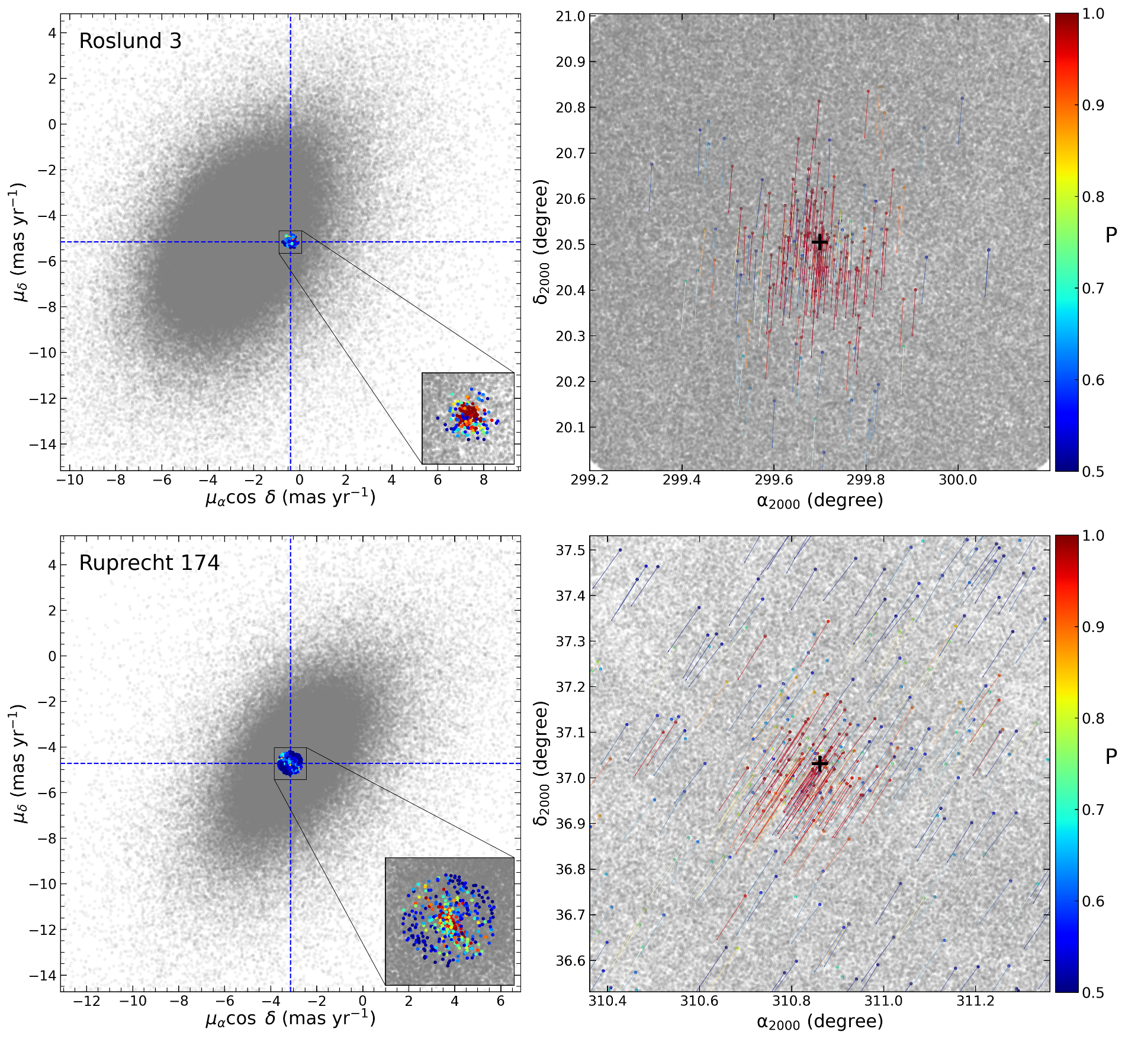}\\
\caption{The fields of Roslund 3 (upper panels) and Ruprecht 174 (lower panels) OCs: {\it Gaia} DR3 proper-motion components (left panels) and sky orientation vectors in equatorial coordinates (right panels). The color scale and circles are consistent with those in Figure~\ref{fig:CMD}. Black plus signs in the right panels mark the central equatorial coordinates of the clusters.}
\label{fig:VPD_all} 
 \end {figure*}

The PM distributions of stars from the \textit{Gaia} catalog within the fields of Roslund 3 and Ruprecht 174, along with their vectorial motion directions in equatorial coordinates, are presented in Figure~\ref{fig:VPD_all}. Analyzing the PM diagrams for the most probable cluster members (colored circles)  confirms that the OCs are separate from the surrounding field stars (gray circles) in Figure~\ref{fig:VPD_all}. However, stars with $P \geq 0.5$ tend to cluster in specific regions, with Roslund 3 displaying a denser stellar population than Ruprecht 174. The most probable members, indicated by colored arrows, show a consistent vectorial alignment in Figure~\ref{fig:VPD_all}. The mean PM components derived from the most probable cluster members were $\langle \mu_{\alpha} \cos \delta, \mu_{\delta} \rangle = (-0.401\pm0.003, -5.169\pm0.003)$ mas yr$^{-1}$ for Roslund 3 and $(-3.139\pm0.006, -4.729\pm0.007)$ mas yr$^{-1}$ for Ruprecht 174. The intersection points of the dashed blue lines in Figure~\ref{fig:VPD_all} correspond to the mean PM values of each cluster.

\begin{figure}[t]
\centering
\includegraphics[width=\columnwidth]{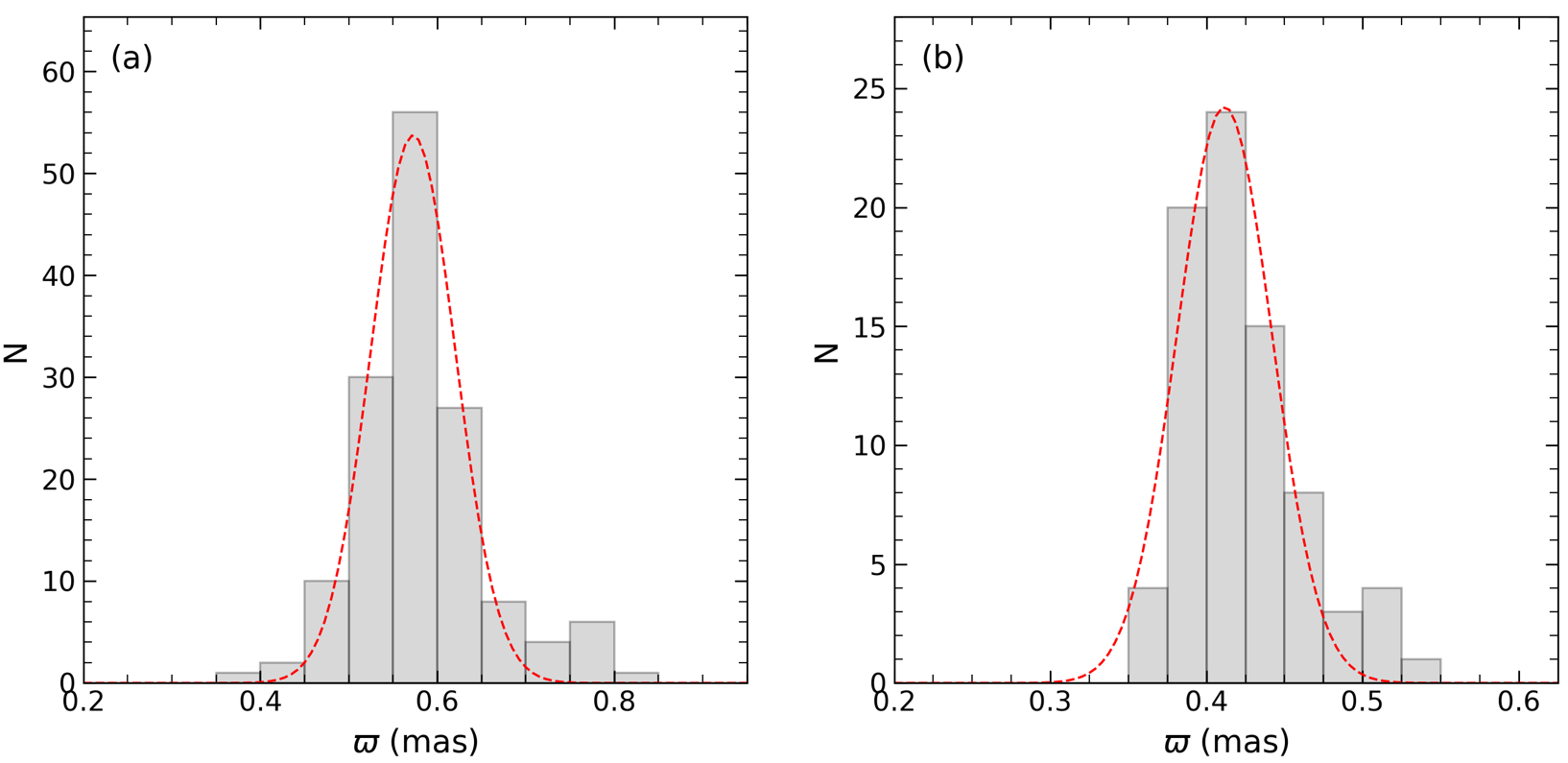}\\
\caption{Trigonometric parallax histograms of stars with membership probabilities $P \geq 0.5$, within the  $G \leq 20.5$ mag limit, and relative parallax errors $\sigma_{\varpi}/\varpi < 0.2$ for the Roslund 3 (a) and Ruprecht 174 (b). The red dashed lines represent the Gaussian curve fitted to the distribution.}
\label{fig:plx_hist}
\end {figure}

The mean trigonometric parallaxes $\langle \varpi \rangle$ of the OCs were estimated by fitting Gaussian functions to the parallax histograms of the members with probabilities $P\geq 0.5$, with a relative parallax error constraint of $\sigma_{\varpi}/\varpi\leq 0.2$ to improve precision. The resulting trigonometric parallaxes are $\langle \varpi \rangle = 0.573\pm 0.004$ mas for Roslund 3 and $0.412 \pm 0.003$ mas for Ruprecht 174. The corresponding trigonometric parallax histograms and Gaussian fits are shown in Figure~\ref{fig:plx_hist}. Using the distance equation $d(\text{pc}) = 1000/\varpi(\text{mas})$, the astrometric distances ($d_{\varpi}$) to the clusters were calculated as $1745 \pm 12$ pc for Roslund 3 and $2427 \pm 18$ pc for Ruprecht 174.

\section{Astrophysical Parameters of the Clusters}
\label{sec:UBV-based}
This section describes the methodologies applied in the astrophysical analysis of two OCs using both \textit{UBV} and \textit{Gaia} photometric data. The \textit{UBV}-based approach involves separate determinations of reddening and photometric metallicities through two-color diagrams (TCDs). Referred to as the \textit{classical method}, this approach is effective in minimizing degeneracy between distance and age. Once these parameters were kept as constants, distance moduli and cluster ages were simultaneously estimated from the CMDs. A detailed explanation of this methodology is available in previous studies \citep{Yontan15, Yontan19, Yontan21, Yontan22, Ak16, Bilir06, Bilir16, Bostanci15, Bostanci18, Koc22, Cakmak24}. Conversely, the determination of all astrophysical parameters in the {\it Gaia} photometric system was carried out simultaneously using the Markov Chain Monte Carlo (MCMC) method.

\subsection{UBV-based Reddenings}
\label{sec:UBV_reddening}
\begin{figure}[h]
\centering
\includegraphics[width=\linewidth]{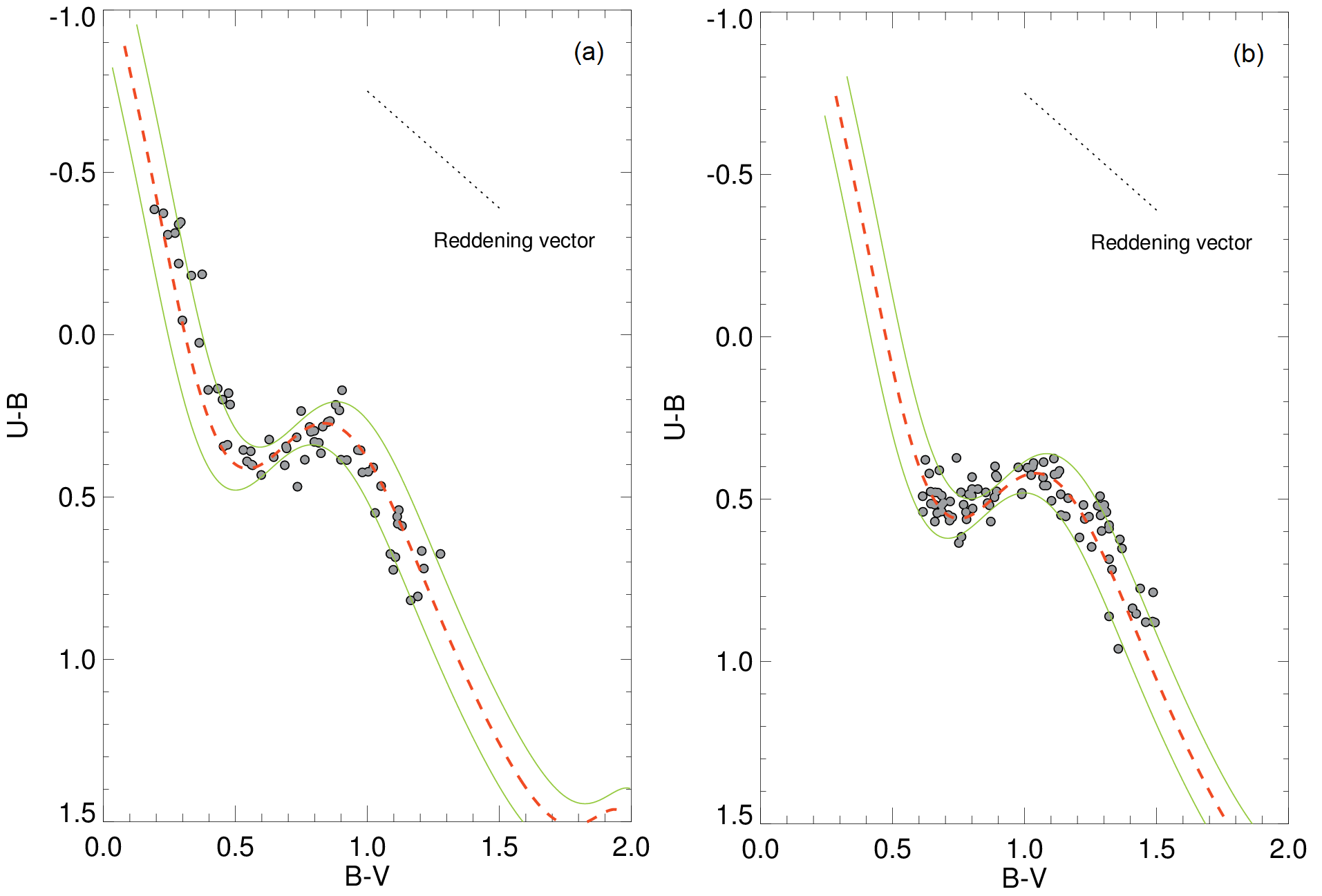}
\caption{The TCDs of the main-sequence members ($P\geq 0.5$) of Roslund 3 (a) and Ruprecht 174 (b). The reddened ZAMS \citep{Sung13}, is represented by the red dashed curves, with the green solid curves indicating $\pm1\sigma$ uncertainty intervals.}
\label{fig:tcds}
\end{figure}

The $U-B$ vs $B-V$ TCDs were employed to estimate the reddening parameters $E(B-V)$ and $E(U-B)$ toward Roslund 3 and Ruprecht 174. To begin, main-sequence stars within the ZAMS curve and with membership probabilities $P\geq 0.5$ were selected and plotted on the TCDs. For Roslund 3, the selected stars span the apparent magnitude range $11 \leq V~{\rm (mag)}\leq 18$, comprising 66 stars, while for Ruprecht 174, the range $13\leq V~{\rm (mag)}\leq 19.5$ includes 90 stars. The observed colors of identified main-sequence stars were compared with the ZAMS for solar metallicity from \citet{Sung13} on TCDs. A $\chi^2$  minimization method was applied, shifting the ZAMS in color space according to the empirical relation $E(U-B)/E(B-V)=0.72+0.05\times E(B-V)$ \citep{Garcia88}, over a range of $0\leq E(B-V)\leq 1.5$ in 0.001 mag increments. The best-fitting values of $E(U-B)$ and $E(B-V)$ were identified as those corresponding to the minimum $\chi^2$ result. Derived reddenings are $E(B-V)=0.410 \pm 0.046$ and $E(U-B)=0.304\pm 0.033$ mag for Roslund 3, and $E(B-V) = 0.615 \pm 0.042$ and $E(U-B)=0.462\pm 0.030$ mag for Ruprecht 174. Figure~\ref{fig:tcds} illustrates the positions of the main-sequence stars with membership probabilities $P\geq 0.5$ on the $U-B$ vs $B-V$ TCDs for Roslund 3 (panel a) and Ruprecht 174 (panel b), overlaid with the best-fit ZAMS curves (dashed red lines).

\subsection{Photometric Metallicity}
\label{sec:metallicity}
To estimate the photometric metallicities [Fe/H] for Roslund 3 and Ruprecht 174, we employed the $UBV$-based technique developed by \citet{Karaali11}, which is designed for F and G-type main-sequence stars within the de-reddened color range $0.3 \leq (B-V)_0~{\rm (mag)} \leq 0.6$. This method relies on calculating the UV-excesses ($\delta$) of cluster members. Initially, we derived the de-reddened color indices $(B-V)_0$ and $(U-B)_0$ for stars within the defined ZAMS curve and with membership probabilities of $P \geq 0.5$, using the $E(B-V)$ and $E(U-B)$ reddening values determined in this study. We applied the color index criterion $0.3 \leq (B-V)_0~{\rm (mag)} \leq 0.6$ \citep{Eker18, Eker20, Eker24, Eker25} to both clusters and identified nine F-G main-sequence stars in total in Roslund 3 and Ruprecht 174. These stars are numbered according to the IDs given in Table~\ref{tab:all_cat}: for Roslund 3, the IDs are 572, 2015, 2058, 2092, 2115, 2174, 2282, 2624 and 3407; and for Ruprecht 174, the IDs are 1035, 1978, 2522, 3453, 3660, 4374, 4402, 4517, 4575 and 4837.
We then constructed $(U-B)_0$ vs $(B-V)_0$ TCDs, along with the Hyades main-sequence data, which served as a reference. This allowed us to calculate the $\delta$ using the relation $\delta = (U-B)_{\rm 0,H} - (U-B)_{\rm 0,S}$, where H and S subscripts represent the Hyades and cluster stars, respectively. Since the calibration reaches its maximum $\delta$ at $(B-V)_0 = 0.6$ mag, the calculated $\delta$ values were normalized using a guillotine factor $f$ \citep{Sandage69}, yielding $\delta_{0.6}$ values for each star, following the approach of \citet{Karaali11}.

\begin{figure*}
\centering
\includegraphics[width=\linewidth]{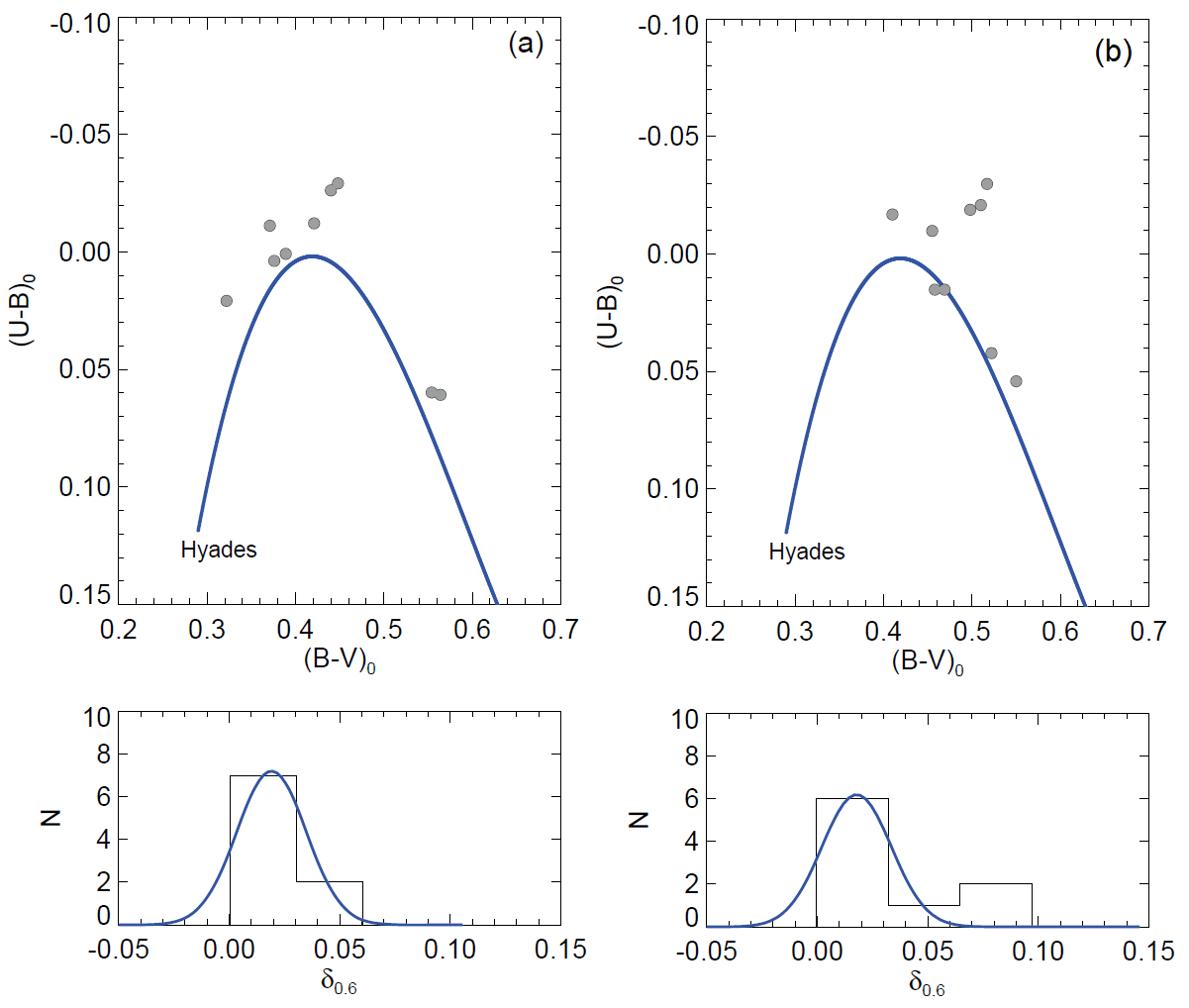}
\caption{$(U-B)_0$ vs $(B-V)_0$ TCDs for nine F–G type main-sequence stars with membership probabilities $P \geq 0.5$ are presented for Roslund~3 (top panel a) and Ruprecht~174 (top panel b). The Hyades main sequence, shown as blue curves, is used as a reference. The bottom panels display the histograms of the normalized $(\delta_{0.6})$ values, along with Gaussian fits represented by blue curves to model the distributions.}
\label{fig:hyades}
\end {figure*}

Next, histograms of the normalized $\delta_{0.6}$ were constructed and fitted with Gaussian functions \citep{Karaali03a, Karaali03b}. The results of the fitting procedure are shown in the lower panels of Figure~\ref{fig:hyades}, and we obtained mean $\delta_{0.6}$ values of $\delta_{0.6}=0.030\pm 0.065$ mag for Roslund 3 and $\delta_{0.6}=0.041\pm 0.064$ mag for Ruprecht 174. The uncertainty in the mean $\delta_{0.6}$ is derived from the standard deviation ($\pm 1\sigma$) of the Gaussian fit. We then used mean $\delta_{0.6}$ values in the photometric metallicity calibration of \citet{Karaali11} given as [Fe/H]~$=-14.316\delta_{0.6}^2-3.557\delta_{0.6}+0.105$.

Consequently, the photometric metallicities for Roslund 3 and Ruprecht 174 were derived as [Fe/H] $= 0.030 \pm 0.065$ and [Fe/H]$= 0.041 \pm 0.064$ dex, respectively. To select appropriate isochrones for the determination of the clusters' ages, the derived [Fe/H] values were converted to mass fraction $Z$ values using the relations recommended by Bovy\footnote{https://github.com/jobovy/isodist/blob/master/isodist/Isochrone.py} for the {\sc parsec} isochrones \citep{Bressan12}. The conversion equations \citep[see also,][]{Gokmen23} as follows:

\begin{equation}
Z_{\rm x}=10\left[{{\rm [Fe/H]}+\log \left(\frac{Z_{\odot}}{1-0.248-2.78\times Z_{\odot}}\right)}\right]
\end{equation}      
and
\begin{equation}
Z=\frac{(Z_{\rm x}-0.2485\times Z_{\rm x})}{(2.78\times Z_{\rm x}+1)}.
\end{equation}
In these formulas, $z_{\rm x}$ is the intermediate variable and $z_{\odot} = 0.0152$ represents the solar metallicity \citep{Bressan12}. From this, we obtained $Z = 0.016$ for Roslund 3 and $Z = 0.017$ for Ruprecht 174.

\subsection{UBV-based Distance Moduli and Age Estimation}
\label{sec:UBV-Distance}
\begin{figure*}
\centering
\includegraphics[width=\linewidth]{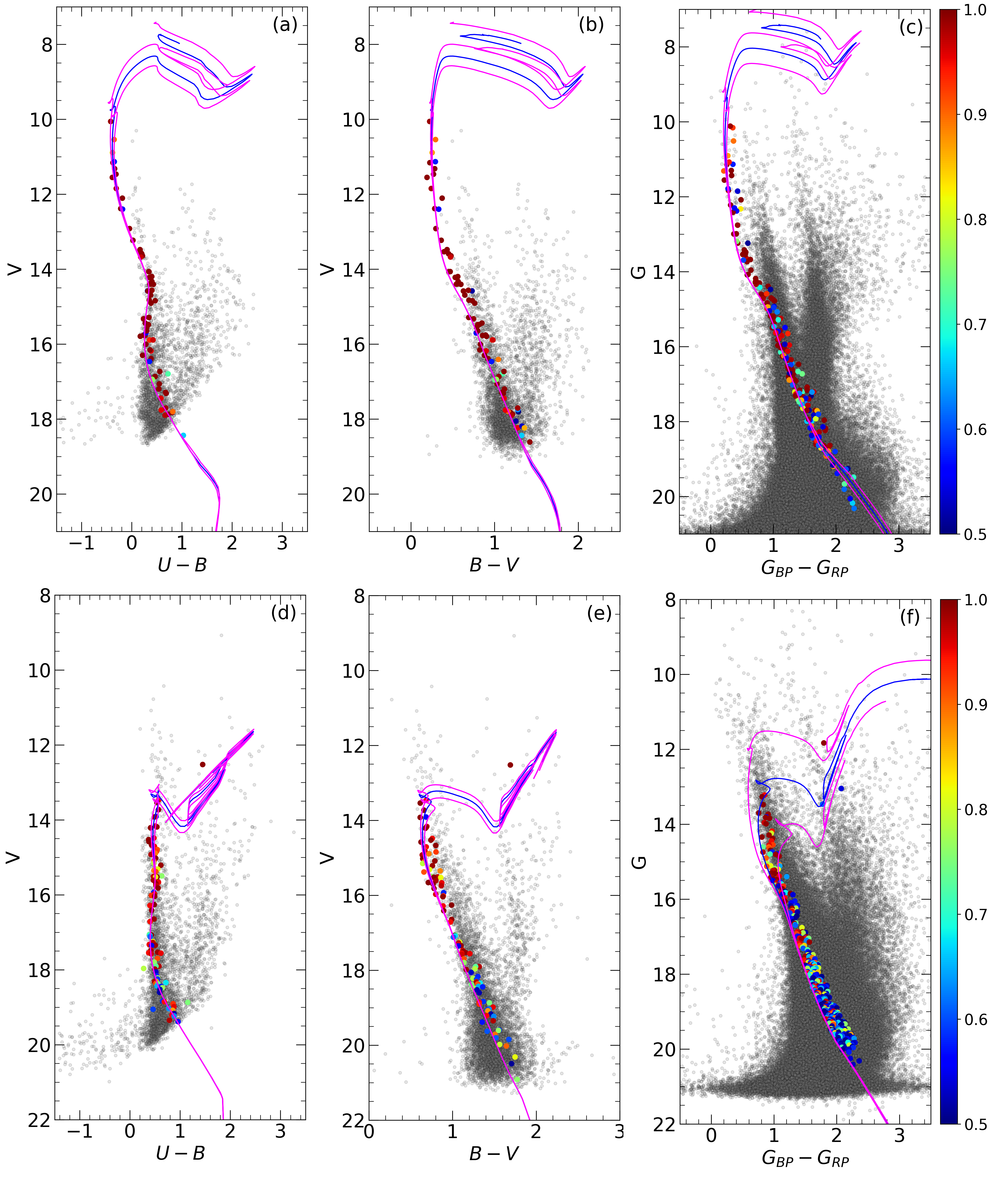}
\caption{CMDs based on $UBV$ and {\it Gaia} photometry for the clusters Roslund 3 (panels a, b, c) and Ruprecht 174 (panels d, e, f). The blue solid lines represent the {\sc parsec} isochrones that best match the observed data, allowing for the determination of the clusters' distance moduli and ages. The purple solid lines indicate the uncertainties associated with the age estimates. The color scales and symbols are consistent with those defined in Figure \ref{fig:CMD}.}
\label{fig:figure_age}
\end {figure*}

The distance moduli ($\mu$) and ages ($\tau$) of Roslund 3 and Ruprecht 174 were derived using the {\sc parsec} isochrones from \citet{Bressan12}. CMDs in the $V$ vs $(U-B)$ and $V$ vs $(B-V)$ planes were created using stars with $P \geq 0.5$ membership probabilities within the cluster regions. The fitting process was performed through visual inspection, focusing on the positions of the most probable member stars ($P \geq 0.5$) in the CMDs. Isochrones of various ages, scaled to the estimated metal fraction $Z$ for each cluster, were overlaid on the $V$ vs $(U-B)$ and $V$ vs $(B-V)$ diagrams. For the $V$ vs $(U-B)$ and $V$ vs $(B-V)$ diagrams, we applied isochrones adjusted for the reddening values $E(U-B)$ and $E(B-V)$, as calculated in Section~\ref{sec:UBV_reddening}. For the distance modulus, we applied equation $(\mu_{\rm V})_{0}=V-M_{\rm V}-A_{\rm V}$, where $V$ is apparent magnitude, $M_{\rm V}$ is the absolute magnitude ,and $A_{\rm V}$ represents the interstellar extinction, calculated as $R\times E(B-V)$. In this study, the total-to-selective extinction ratio $R$ is adopted as 3.1 \citep{Cardelli89}. The isochrone fitting process considered the main-sequence, turn-off, and giant stars with $P \geq 0.5$ for each cluster. Distance moduli and distance errors were derived using the methodology of \citet{Carraro17}, while age uncertainties were estimated by fitting low and high-age isochrones to the observed scatter in the main sequence and turn-off. The best-fit isochrones are shown in Figure~\ref{fig:figure_age} for the $V$ vs $(U-B)$ and $V$ vs $(B-V)$ CMDs.

As shown in Figure~\ref{fig:figure_age}, the {\sc PARSEC} isochrones \citep{Bressan12} fitted to the cluster member stars generally provide a good representation of the clusters’ CMD morphology. However, this agreement does not hold for the two giant stars in Ruprecht 174. The discrepancy between the positions of these two stars, both of which have high membership probabilities, and the {\sc PARSEC} isochrones may be attributed to several factors: they could be unresolved binary systems, their metallicities might differ from the cluster's mean value, or they may exhibit additional extinction due to circumstellar dust, as they are particularly cold stars. To clarify the nature of this inconsistency, more detailed spectroscopic observations are required.

The results derived from the isochrone fitting to the CMDs are as follows: For Roslund 3, isochrones with ages $\log \tau~({\rm yr})=7.70$, $7.78$, and $7.85$ and a metal fraction of $z=0.016$ were superimposed onto the {\it UBV}-based CMDs. This resulted in an apparent distance modulus of $(\mu_{\rm V})_0=12.407\pm 0.150$ mag, which corresponds to an isochrone-based distance of $d_{\rm iso}=1687\pm 121$ pc. The main-sequence and turn-off member stars were taking into account by visual inspection, leading to an estimated cluster age of $\tau=60\pm 6$ Myr. For Ruprecht 174, isochrones with ages $\log \tau~({\rm yr}) = 8.67$, $8.72$, and $8.76$ and a metal fraction of $z=0.017$ were applied. The calculated distance modulus is $(\mu_{\rm V})_{0}=13.794\pm 0.144$ mag, corresponding to an isochrone-based distance of $d_{\rm iso}=2385\pm 163$ pc. The cluster’s age was determined to be $\tau=520\pm 50$ Myr. 

\begin{figure}
\centering
\includegraphics[width=\linewidth]{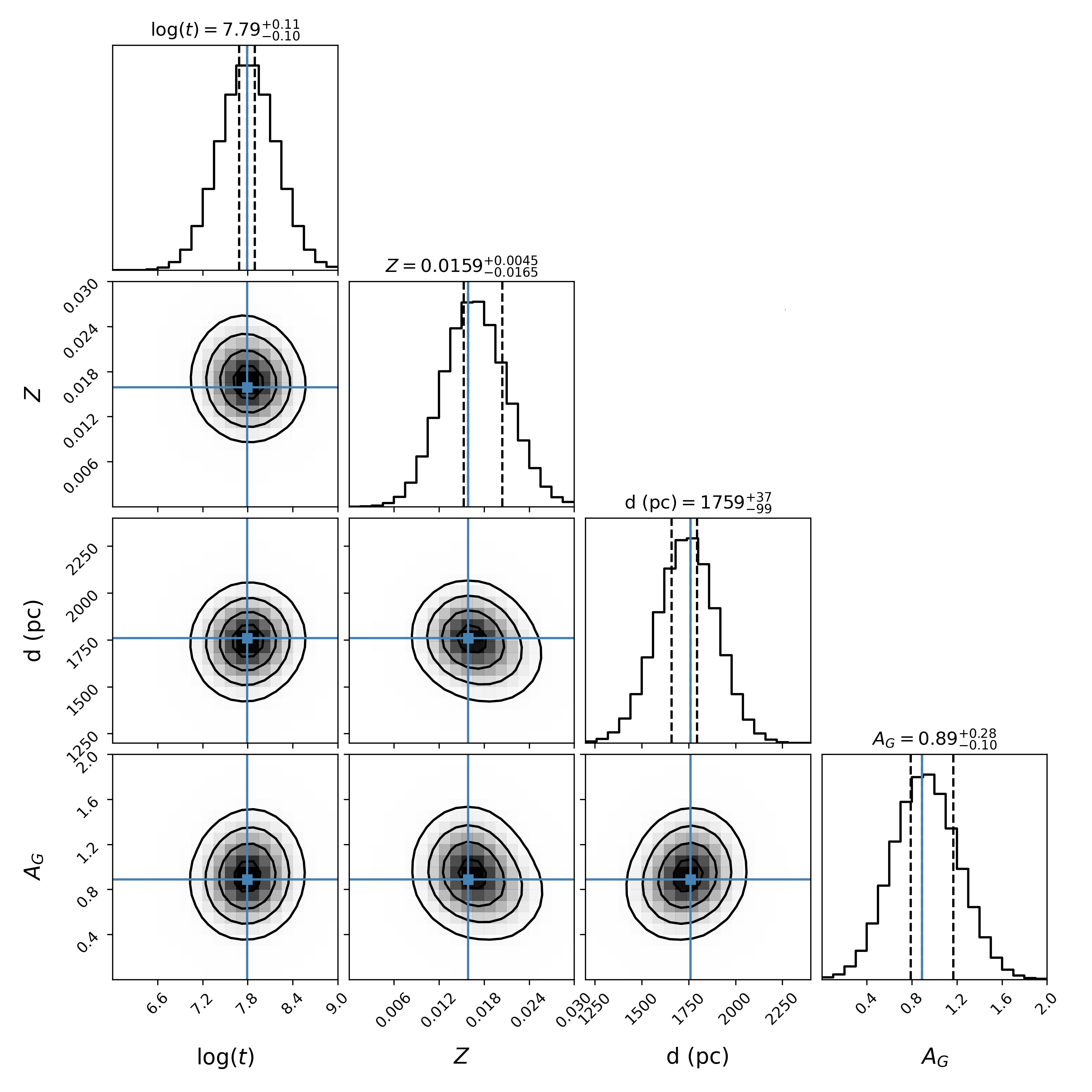}\\
\includegraphics[width=\linewidth]{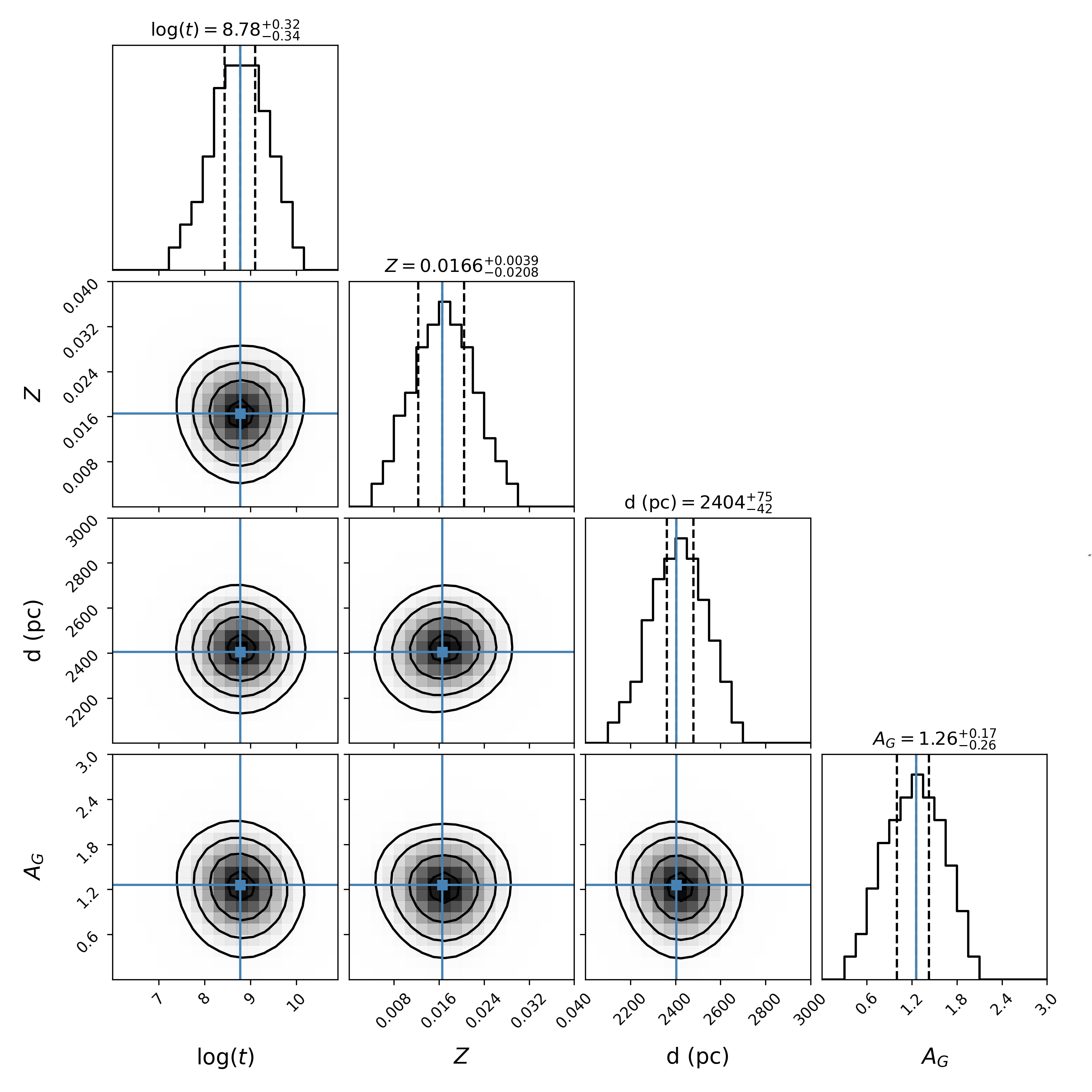}\\
\caption{The corner plot presents both the one-dimensional marginalized and two-dimensional joint posterior probability distributions for the inferred parameters of Roslund~3 and Ruprecht~174. In the two-dimensional panels, the overlaid contours correspond to confidence levels of 68\%, 90\%, and 95\%, delineating the regions of highest probability density. The one-dimensional histograms display the posterior distributions for each parameter individually, where the vertical lines mark the median values along with the 16th and 84th percentiles, representing the 1$\sigma$ credible intervals.}
\label{fig:figure_ten} 
\end {figure}

\subsection{\textit{Gaia}-based Astrophysical Parameters}
In the \textit{classical method} (see Section~\ref{sec:UBV-Distance}) the astrophysical parameters were derived individually. In contrast, within the {\it Gaia} photometric system, the MCMC technique was employed to simultaneously estimate the reddening, metallicity, distance, and age. This analysis utilized stars with membership probabilities of $P \geq 0.5$, corresponding to 198 likely members for Roslund 3 and 397 for Ruprecht 174.

\subsubsection{Markov-Chain Monte Carlo Analyses}
\label{sec:mcmc}

We developed a method grounded in the Monte Carlo approach to independently estimate the distance ($d$), extinction in the $G$ band ($A_{\rm G}$), age ($\tau$), and metal fraction ($Z$) of an OC. In this approach, we set the number of MCMC walkers equal to the number of probable cluster member stars. Each walker is assigned to one star and evaluates how well that star’s observed properties match synthetic values from the isochrone grid, assuming a common set of cluster parameters: distance, extinction, age, and metallicity. While the walkers explore the parameter space independently based on their assigned star, they all sample the same global parameters, which represent the cluster as a whole. The joint posterior distribution is then constructed by combining the likelihood contributions from all walkers. This design allows us to incorporate the information from all stars while converging on a single best-fitting solution for the cluster. The isochrone grid was constructed using version CMD 3.8 of the {\sc parsec} stellar evolutionary models \citep{Bressan12}. This grid spans the age range $6 \leq \log \tau~{\rm (yr)} \leq 10.13$ in steps of 0.05 in $\log \tau$, and covers metal fraction values from $Z=0$ to $Z=0.03$ in increments of 0.0005 dex.

The Monte Carlo sampling was performed over a parameter space that included distances from 0 to 30 kpc and extinction values ($A_{\rm G}$) between 0 and 10 magnitudes, while age and metallicity were sampled across the full extent of the defined isochrone grid. To achieve a continuous parameter space and minimize the influence of the discrete step sizes in the input grid, we employed the Delaunay interpolation function from the \texttt{Scipy} Python library \citep{Scipy}. This three-dimensional interpolation technique allows for smooth transitions across the grid, ensuring a more robust and precise application of the Monte Carlo method.

To investigate the posterior distributions of the cluster parameters, we applied a Monte Carlo sampling technique using the affine-invariant Markov Chain Monte Carlo (MCMC) algorithm implemented in the Python \texttt{emcee} library \citep{emcee}. For each cluster, the number of walkers was set equal to the number of probable member stars, and each walker was evolved over 5,000 iterations. The final estimates for the parameters were derived from the resulting posterior distributions by extracting the 16th, 50th, and 84th percentiles, providing robust uncertainty estimates. This method enables each star to independently explore the parameter space while collectively converging on a global solution representative of the cluster, thereby facilitating the simultaneous determination of distance, extinction, age, and metallicity.

As a result of this analysis, we obtained estimates for the ages, heavy-element abundances, distances, and $G$-band extinction coefficients ($A_{\rm G}$) of the two OCs. The outcomes are listed in Table~\ref{tab:07}. The posterior distributions of $A_{\rm G}$, $d$, $Z$, and $\log\tau$ are shown in Figure~\ref{fig:figure_ten}, where the two-dimensional marginalized distributions include contours representing confidence intervals of \(68\%\), \(90\%\), and \(95\%\). The one-dimensional projections show vertical dashed lines marking the 16th, 50th, and 84th percentiles. Based on the MCMC-derived parameters from the {\it Gaia} photometric data, we constructed CMDs and overplotted the corresponding {\sc parsec} isochrones. As seen in Figures~\ref{fig:figure_age}c and f, the modeled isochrones align well with the observed {\it Gaia}-based CMDs, confirming the consistency and reliability of the parameter estimates.

\begin{table}[t]
  \centering
  \caption{Fundamental parameters of the Roslund 3 and Ruprecht 174 derived using the MCMC for {\it Gaia} DR3 data.}
    \begin{tabular}{lcc}
    \hline
    & Roslund 3 & Ruprecht 174 \\
    \hline
    Parameter          & $G$ vs $(G_{\rm BP}-G_{\rm RP}$) & $G$ vs $(G_{\rm BP}-G_{\rm RP}$) \\
    \hline
    $A_{\rm G}$ (mag)  & $0.89^{+0.28}_{-0.10}$           & $1.26^{+0.17}_{-0.26}$       \\
    $d$         (pc)   & $1759^{+37}_{-99}$               & $2404^{+75}_{-42}$           \\
    $Z$                & $0.0159^{+0.0045}_{-0.0165}$     & $0.0166^{+0.0039}_{-0.0208}$ \\
    $\log \tau$ (yr)   & $7.79^{+0.11}_{-0.10}$           & $8.78^{+0.32}_{-0.34}$       \\
    \hline
    \end{tabular}%
  \label{tab:07}%
\end{table}%

\section{Orbital Analyses}
Radial velocity ($V_{\rm R}$) measurements are needed for kinematic and orbital dynamic analyses of OCs. In this study, the mean radial velocities of Roslund 3 and Ruprecht 174 were determined using $V_{\rm R}$ data obtained from the {\it Gaia}-based catalog. Only stars with cluster membership probabilities of $P \geq 0.5$ were considered in the analysis. The {\it Gaia} DR3 provided radial velocity measurements for nine stars in Roslund 3 and four stars in Ruprecht 174. The stars used in the calculation of the mean radial velocities of two clusters are listed in Table~\ref{tab:RV_info} along with their {\it Gaia} DR3 identifiers. Mean radial velocities ($\langle V_{\rm R}\rangle$) for each cluster were computed using weighted averages, following the methodology described by \citet{Soubiran18}. The results yielded a $\langle V_{\rm R}\rangle$ of $-15.42\pm 7.52$ km s$^{-1}$ for Roslund 3 and  $-11.99\pm 2.5$ km s$^{-1}$ for Ruprecht 174. Space velocity components and Galactic orbital parameters of the clusters were computed using the {\sc MWPotential2014} model within the {\tt galpy} library \citep{Bovy15}. For Roslund 3 and Ruprecht 174, the input parameters employed in the space velocity and orbit calculations including equatorial coordinates ($\alpha$, $\delta$), $\langle V_{\rm R}\rangle$ obtained in this study, mean PM components ($\mu_{ \alpha}\cos\delta, \mu_{\rm \delta}$), distances ($d_{\rm iso}$) derived from main-sequence fitting, and their associated uncertainties. The adopted parameters for these calculations were a Galactocentric distance of $R_{\rm gc} = 8$ kpc, a rotation velocity of $V_{\rm rot} = 220$ km~s$^{-1}$ \citep{Bovy15, Bovy12}, and a vertical displacement of the Sun from the Galactic plane of $27 \pm 4$ pc \citep{Chen00}. The past orbital trajectories of Roslund 3 and Ruprecht 174 were integrated backward over a timespan of 2.5 Gyr with a temporal resolution of 1 Myr. To enable an accurate and precise determination of their Galactic orbital parameters, the analysis was extended until each cluster completed a well-defined closed orbit. The analysis provided key orbital parameters, including the Galactocentric distance ($R_{\rm gc}$), apogalactic ($R_{\rm a}$), and perigalactic ($R_{\rm p}$) distances, orbital eccentricities ($e$), maximum vertical deviations from the Galactic plane ($Z_{\rm max}$), space velocity components ($U$, $V$, $W$), and orbital periods ($P_{\rm orb}$).

\begin{figure*}
\centering
\includegraphics[width=\linewidth]{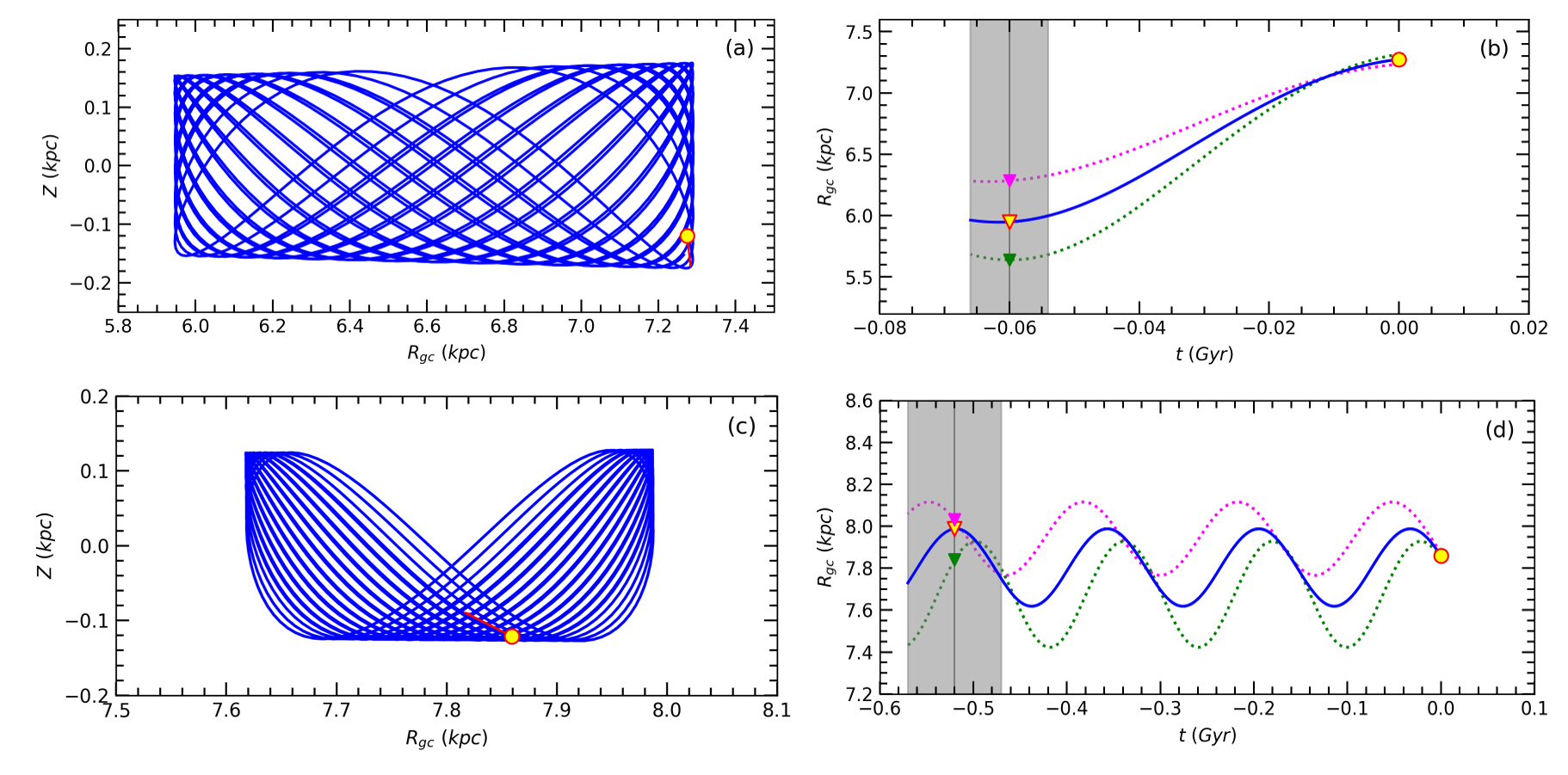}
\caption{The orbital motions of the Roslund 3 (a, b) and Ruprecht 174 (c, d) in the $(Z$ vs $R_{\rm {gc}})$ and $(R_{\rm {gc}}$ vs $t)$ planes.} The filled yellow circles and triangles represent the present-day and early orbital positions of the clusters in the Galaxy, respectively. Red arrows indicate the direction of their motion.
\label{fig:galactic_orbits}
\end {figure*}

\begin{table}
  \centering
\setlength{\tabcolsep}{2.5pt}
\renewcommand{\arraystretch}{0.8}
\caption{\textit{Gaia} DR3 names, equatorial coordinates, radial velocities and their errors of member stars used in the calculation of mean radial velocities for both clusters.}
{\scriptsize
    \begin{tabular}{cccr}
      \hline
  Gaia DR3 Name & RA & DEC & $V_{\rm R}$~~~~~~~~\\
  &(hh:mm:ss.ss) &	(dd:mm:ss.ss)& (km s$^{-1}$)~~~~~\\
  \hline
    \multicolumn{4}{c}{Roslund 3}  \\
  \hline  
  1823381331371670400 &19:57:53.15 & +20:27:56.34 & $12.75\pm18.15$  \\
  1823386725851173120 &19:58:14.74 & +20:29:06.54 & $-0.60 \pm 6.92$ \\
  1823373840948393600 &19:58:33.95 & +20:26:31.19 & $23.42\pm13.00$  \\
  1823390093105528064 &19:58:35.35 & +20:36:54.89 & $-24.75\pm12.54$ \\  
  1823376654120364032 &19:58:42.82 & +20:31:29.14 & $-17.84\pm15.19$ \\
  1823376864591083904 &19:58:49.84 & +20:31:29.90 &  $3.50\pm15.35$  \\
  1823308591801358720 &19:58:50.14 & +20:02:39.15 & $-24.13\pm21.16$ \\
  1823376108691033856 &19:59:02.56 & +20:31:06.38 & $-51.71\pm17.71$ \\
  1823375662014401536 &19:59:01.75 & +20:28:15.04 & $-49.23\pm8.22$  \\
  \hline
  \multicolumn{4}{c}{Ruprecht 174} \\
  \hline
  1870648649031735296 & 20:42:59.23 & +36:30:07.47 & $-22.61\pm1.01$  \\
  1870681050266327552 & 20:43:18.23 & +37:02:31.96 & $192.20\pm12.93$ \\
  1870677850499835392 & 20:43:27.15 & +37:00:49.09 &  $-6.81\pm0.33$  \\
  1870604977802400000 & 20:44:13.89 & +36:31:05.36 & $-15.67\pm0.31$  \\
      \hline
    \end{tabular}%
  \label{tab:RV_info}%
  }
\end{table}%

To investigate the three-dimensional locations of both OCs, we determined their projected Cartesian coordinates in the Galactic reference frame based on the heliocentric distances of their member stars. These positions, denoted as $(X, Y, Z)_{\odot}$, are calculated using the following equations: $X = d_{\rm iso} \cos b \cos l$, $Y=d_{\rm iso} \cos b \sin l$, and $Z=d_{\rm iso}\sin b$ is the heliocentric distance, and $(l, b)$ are the Galactic longitude and latitude of the OC center, respectively. $X$ denotes the distance in the direction toward the Galactic center, $Y$ represents the distance along the direction of the Sun’s rotation, and $Z$ corresponds to the distance from the Galactic plane toward the North Galactic Pole. The analyses indicate that the Cartesian coordinates for Roslund 3 are $(X,Y,Z)_{\odot}=(870, 1439, 138)$ pc, while those for Ruprecht 174 are $(X, Y, Z) _{\odot}= (494, 2329, 141)$ pc (see Table~\ref{tab:Final_table}). The distances of the clusters from the Galactic center were calculated using the relation $R_{\rm gc} = \sqrt{R_{\rm 0}^{2} + (r \cos b)^2 - 2 R_{\rm 0} r \cos b \cos l}$ \citep{Guctekin19, Cinar2025}. Here, $R_{\rm 0}=8$ kpc denotes the Sun’s distance from the Galactic center \citep{Majewski93}. The results show that the Galactocentric distances of Roslund 3 and Ruprecht 174 are 7270$\pm$24 and 7860$\pm$15 pc, respectively (see Table~\ref{tab:Final_table}).

This study introduces the concept of the traceback early orbital radius ($R_{\rm teo}$), as defined by \citet{Akbaba24}, which characterizes the past orbital configuration of a stellar system by integrating its orbit under the assumption of a static Galactic potential. Unlike the birth radius ($R_{\rm birth}$), which aims to identify the exact location where a cluster formed, $R_{\rm teo}$ offers insight into its earlier orbital position within the Galaxy. The calculation of $R_{\rm teo}$ was carried out alongside other kinematic and dynamical parameters. The results, summarized in Table~\ref{tab:Final_table}, indicate that the space-velocity components ($U$, $V$, $W$) of Roslund 3 are $(22.91 \pm 6.06, -33.60 \pm 4.96, -17.40 \pm 1.96)$ km s$^{-1}$, whereas those of Ruprecht 174 are $(60.03 \pm 4.69, -25.28 \pm 1.54, -4.32 \pm 0.50)$ km~s$^{-1}$. It is important to note that these velocities are heliocentric and therefore include the solar motion. To eliminate this bias, Local Standard of Rest (LSR) corrections were applied using the solar motion values provided by \citet{Coskunoglu11}. The corrected space velocity components $(U, V, W)_{\rm LSR}$, along with the total space velocity $S_{\rm LSR}$, are also listed in Table~\ref{tab:Final_table}.

As illustrated in Figure~\ref{fig:galactic_orbits}, the Galactic orbits of Roslund 3 and Ruprecht 174 are shown. Panels (a) and (c) illustrate the motion of the clusters within the Galaxy in terms of $R_{\rm gc}$ and $Z$ \citep[e.g.][]{Tasdemir23, Tasdemir2025, TasdemirCinar2025}. Panels (b) and (d) show how the Galactocentric distances of the clusters evolve over time with $R_{\rm gc}$ vs $t$ plane \citep[e.g.][]{Yontan23c, Yucel24, Yucel25, Haroon2025}. In Figure~\ref{fig:galactic_orbits}, the current and early orbital positions of Roslund 3 and Ruprecht 174 are indicated by yellow-filled circles and triangles, respectively. Dashed pink and green lines, along with their associated triangles, represent the orbital trajectories and early orbital positions of the clusters, incorporating uncertainties in the input parameters.

The upper panels of Figure~\ref{fig:galactic_orbits} show that both of the OCs originated within the Solar circle and have remained within this region throughout their orbit. The uncertainties in the age estimates of the clusters significantly influence the determination of their early orbital positions. The age uncertainties were found to be 6 Myr for Roslund 3 and 50 Myr for Ruprecht 174. These uncertainties were incorporated into the analysis, with the variation in the $R_{\rm teo}$ positions of both clusters depicted as shaded regions in the right panels of Figure~\ref{fig:galactic_orbits}. Dynamic orbital analyses indicate that the Roslund 3 and Ruprecht 174 OCs birthplaces were at distances of $5.95\pm0.32$ and $7.99\pm0.10$ kpc, respectively, from the Galactic center (see also, Table~\ref{tab:Final_table}).

The derived orbital parameters of the two OCs were utilized to infer their Galactic population classifications. Based on the criteria proposed by \citet{Schuster12}, stars with $V_{\rm LSR}$ velocities in the range $V_{\rm LSR}~({\rm km~s^{-1}) } > -50$ are associated with the thin disk, those within $-180 < V_{\rm LSR}~({\rm km~s^{-1}}) \leq -50$ belong to the thick disk, and values below $-180$ km s$^{-1}$ correspond to halo membership. Given the calculated $V_{\rm LSR}$ values for Roslund 3 and Ruprecht 174, both OCs are identified as members of the Galactic thin disk. Their orbital motions are characterized by low eccentricities (not exceeding 0.1) and vertical excursions from the Galactic plane limited to $Z_{\rm max}=175\pm26$ pc for Roslund 3 and $Z_{\rm max}=128\pm9$ pc for Ruprecht 174. These kinematic features further support their classification as thin-disk clusters \citep{Plevne15, Guctekin19}.


\section{Investigating the Dynamical Behavior of the Clusters}
\subsection{Luminosity Function}
\begin{figure}
\centering
\includegraphics[width=\linewidth]{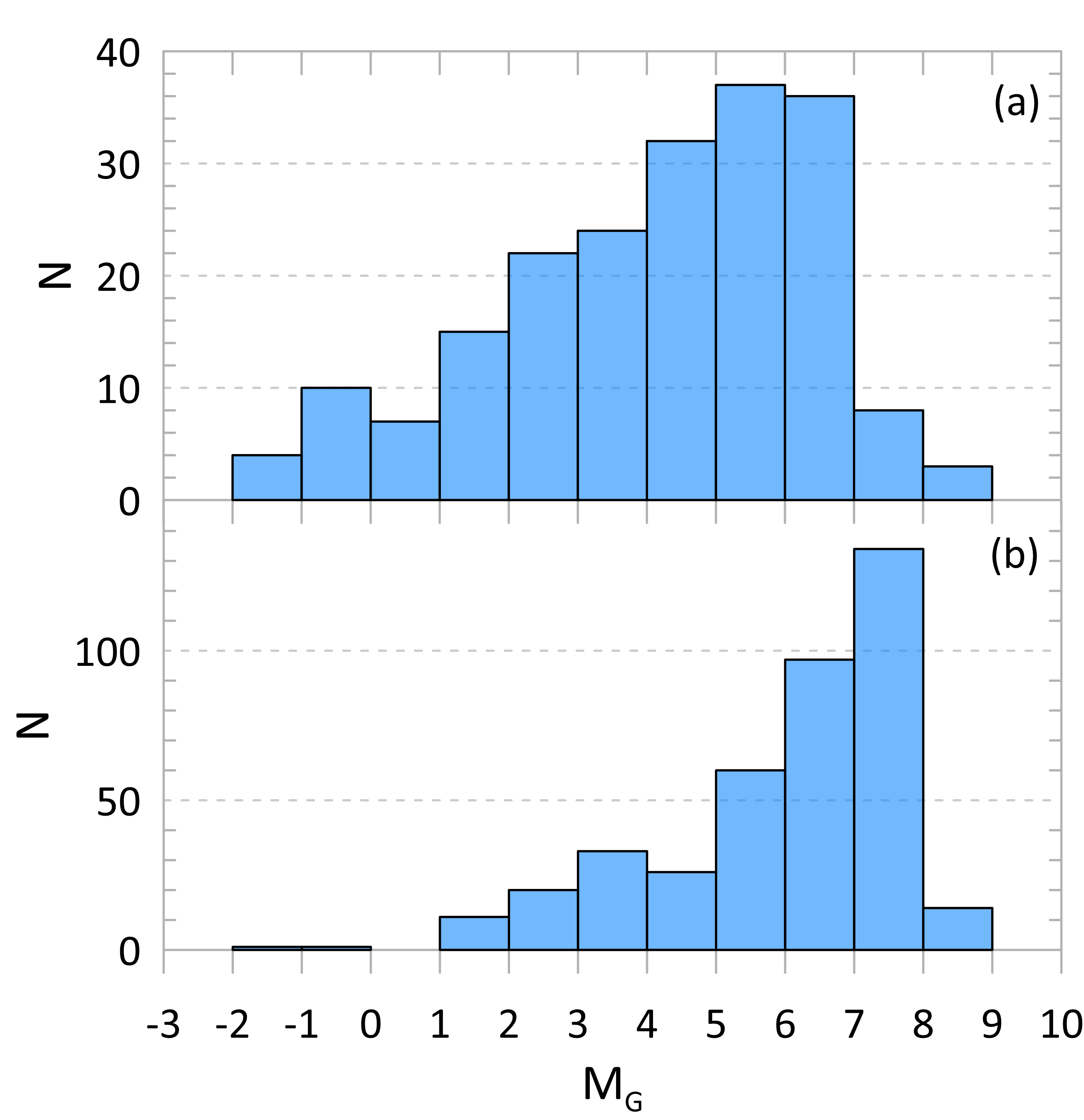}
\caption{LF histograms for Roslund 3 (a) and Ruprecht 174 (b).}
\label{fig:luminosity_functions}
\end {figure}

The luminosity functions (LFs) of both OCs are derived based on the main-sequence stars of each cluster in this study. To determine the luminosity functions (LFs) of the OCs, stars with membership probabilities $P\geq0.5$ were selected using astrometric and photometric data from {\it Gaia} DR3. Main-sequence stars with apparent magnitudes in the range $10 \leq G~{\rm (mag)} \leq 20.5$ for Roslund 3 and $13 \leq G~{\rm (mag)} \leq 20.5$ for Ruprecht 174 were used in the analyses. The absolute magnitudes, $M_{\rm G}$, of these stars were computed using the distance modulus equation of $M_{\rm G} = G - 5\times\log d_{\rm iso}+5-A_{\rm G}$. The extinction value in the $G$-band ($A_{\rm G}$) is also calculated using the relation $A_{\rm G} = R\times E(G_{\rm BP} - G_{\rm RP})$, where $R$ represents the selective absorption coefficient specific to the {\it Gaia} the $G$-band. A value of $R = 1.8626$ was adopted based on the studies of \citet{Cardelli89} and \citet{ODonnell94}. The resulting absolute magnitude ranges were found to be $-1.85 < M_{\rm G}~({\rm mag}) < 8.35$ for Roslund 3 and $1.15 < M_{\rm G}~({\rm mag})< 8.53$ for Ruprecht 174. Consequently, histograms of the LFs were constructed by counting the number of stars within unit intervals of $M_{\rm G}$ (Figure~\ref{fig:luminosity_functions}). As shown in Figure~\ref{fig:luminosity_functions}, the stellar populations in both Roslund 3 (a) and Ruprecht 174 (b) display an increasing trend up to approximately $M_{\rm G} \sim 8$ mag, followed by a decline beyond this magnitude limit.

\subsection{Present Day Mass Function}
\begin{figure}
\centering
\includegraphics[width=\linewidth]{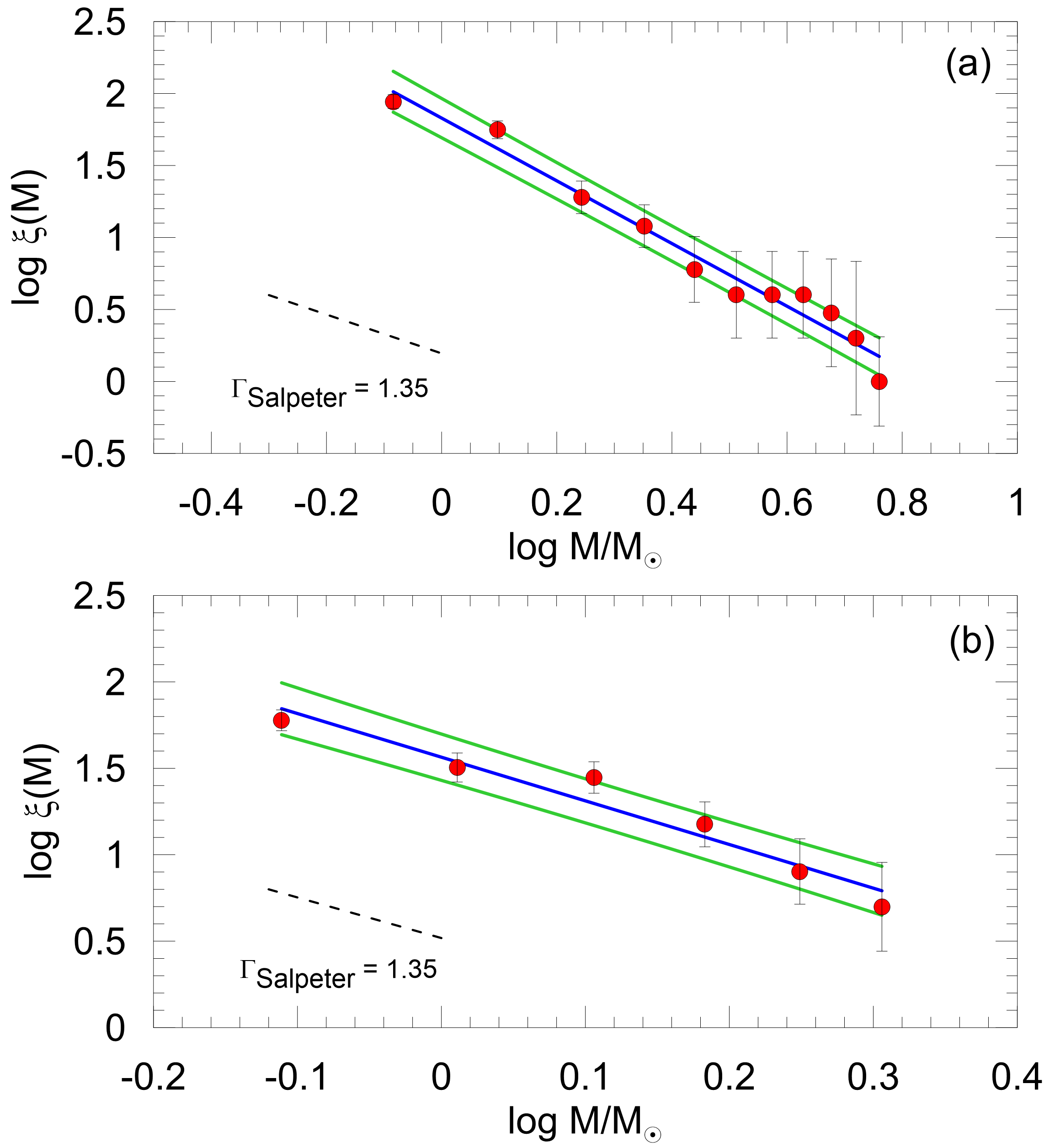}
\caption{PDMFs for Roslund 3 (a) and Ruprecht 174 (b). The red points represent the mass distributions, while the blue line shows the best-fit linear trend. Green lines mark the one standard deviation boundaries of the fit.}
\label{fig:mass_functions}
\end {figure}

The present-day mass functions (PDMFs) of both OCs were also determined. Stellar masses were estimated using theoretical {\sc PARSEC} models, which were previously employed to derive the clusters’ ages and distance moduli \citep{Bressan12}. These {\sc PARSEC} models provide absolute magnitudes and stellar masses across multiple photometric systems. To establish a correlation between absolute magnitude and stellar mass, a polynomial function was fitted to the data. Separate mass-luminosity relations were derived for each cluster, enabling the calculation of main-sequence stellar masses identified in the LF analysis. As a result, the mass ranges of main-sequence stars were found to be $0.5 < M/M_{\odot} \leq 5.8$ for Roslund 3 and $0.5 < M/M_{\odot} \leq 2.2$ for Ruprecht 174. To estimate the PDMFs, stellar mass distributions were generated by grouping stars into mass bins of width $0.5~M/M_{\odot}$ and $0.25~M/M_{\odot}$ for Roslund 3 and Ruprecht 174, respectively. The logarithmic values of the number of stars per mass bin were then calculated, and the resulting distributions are shown in Figure~\ref{fig:mass_functions}. Uncertainties in the stellar mass values were estimated using Poisson statistics. The PDMFs of the clusters were derived by applying the linear relationship $\log \left( \frac{dN}{dM} \right) = -(1 + \Gamma) \times \log M + \text{constant}$; where $dN$ is the number of stars per unit mass interval, $M$ is the central mass of the stars, and $\Gamma$ is the PDMF's slope, as described by \citet{Salpeter55}. This relationship is shown by the solid blue line in Figure~\ref{fig:mass_functions}. The resulting PDMF slopes for Roslund 3 and Ruprecht 174 are determined to be $\Gamma = 1.18 \pm 0.13$ and $\Gamma = 1.53 \pm 0.30$, respectively. These findings are consistent with the canonical slope of $\Gamma = 1.35$, as proposed by \citet{Salpeter55}.

\subsection{The Dynamical State of Mass Segregation}

Mass segregation is a dynamical process in which more massive stars progressively migrate toward the central regions of a star cluster due to gravitational interactions, while lower-mass stars are displaced outward \citep{Sagar88, Raboud98, Fischer98}. This phenomenon, driven by long-term relaxation mechanisms, plays a central role in shaping the internal structure and dynamical evolution of clusters \citep{Heggie79, Hillenbrand98, Baumgardt03, Dib19, Bisht20, Pavlik22}. A key parameter in quantifying this effect is the half-mass relaxation time ($T_{\rm E}$), which estimates the timescale over which stellar encounters redistribute kinetic energy, leading toward a state of approximate energy equipartition and a Maxwellian velocity distribution. The relaxation time depends on the total number of stars ($N$), the half-mass radius ($R_{\rm h}$), and the mean stellar mass ($\langle m \rangle$), and can be approximated by the relation from \citet{Spitzer71}:
\begin{equation}
T_{\rm E} = \frac{8.9 \times 10^{5} N^{1/2} R_{\rm h}^{3/2}}{\langle m\rangle^{1/2}\log(0.4N)}.
\end{equation} 
Beyond internal dynamics, the extent of mass segregation also varies with the cluster's age and its interaction with external agents such as tidal fields and stellar feedback \citep{Kruijssen12}. In early evolutionary stages, clusters may exhibit primordial mass segregation before reaching dynamical equilibrium. Over time, relaxation enhances mass stratification, particularly in dense environments. Studies show that young or dynamically evolved clusters with shorter relaxation times tend to display strong central concentrations of massive stars \citep{Allison09, Gieles11}, whereas older systems often show more uniform mass distributions due to prolonged mixing.

In this study, we estimated the relaxation times for both of the analyzed OCs and evaluated these values concerning the extent of the observed mass segregation. The stellar sample used in the analysis consisted of stars with membership probabilities of $P\geq 0.5$, all located within the limiting radius of each cluster, which was defined as 40 arcminutes. Although the $r_{\rm lim}$ of the clusters were determined from the stellar density profiles, a fixed radius of 40 arcminutes was adopted to ensure a statistically significant stellar sample, particularly for the investigation of mass segregation in the cluster halos. In some cases, the number of stars within the derived limiting radius was insufficient to reliably assess the dynamical structure. As a result, 198 members were identified in Roslund 3 and 397 in Ruprecht 174, spanning mass ranges of $0.5 < M/M_{\odot} \leq 5.8$ and $0.5 < M/M_{\odot} \leq 2.2$, respectively. The total masses were estimated as $289~M_{\odot}$ for Roslund 3 and $354~M_{\odot}$ for Ruprecht 174, corresponding to mean stellar masses of $\langle m \rangle = 1.46~M_{\odot}$ and $0.89~M_{\odot}$, respectively. These mean stellar masses were derived by dividing the total cluster mass by the number of member stars included in the mass determination. This approach provides a representative average mass per star within the analyzed stellar samples of each cluster. The derived half-mass radii were $R_{\rm h} = 2.65$ and $9.83$ pc, leading to relaxation times of about $T_{\rm E} = 24$ Myr and $T_{\rm E} = 263$ Myr, respectively. Since the calculated $T_{\rm E}$ values for the two OCs are shorter than the estimated cluster ages (see Table~\ref{tab:Final_table}), we conclude that Roslund 3 and Ruprecht 174 have reached a dynamically relaxed state. It is important to note that the relaxation time estimates provided in this study represent lower limits. These values are derived exclusively from stars detected within the photometric completeness threshold. Undetected low-mass stars and unresolved binary systems may contribute additional mass and increase the total stellar count, potentially leading to longer relaxation times than currently estimated. In the case of Roslund 3, this suggests that the actual relaxation time might exceed the cluster’s age, indicating that the system may not yet have reached full dynamical equilibrium.

\begin{figure}
\centering
\includegraphics[width=\linewidth]{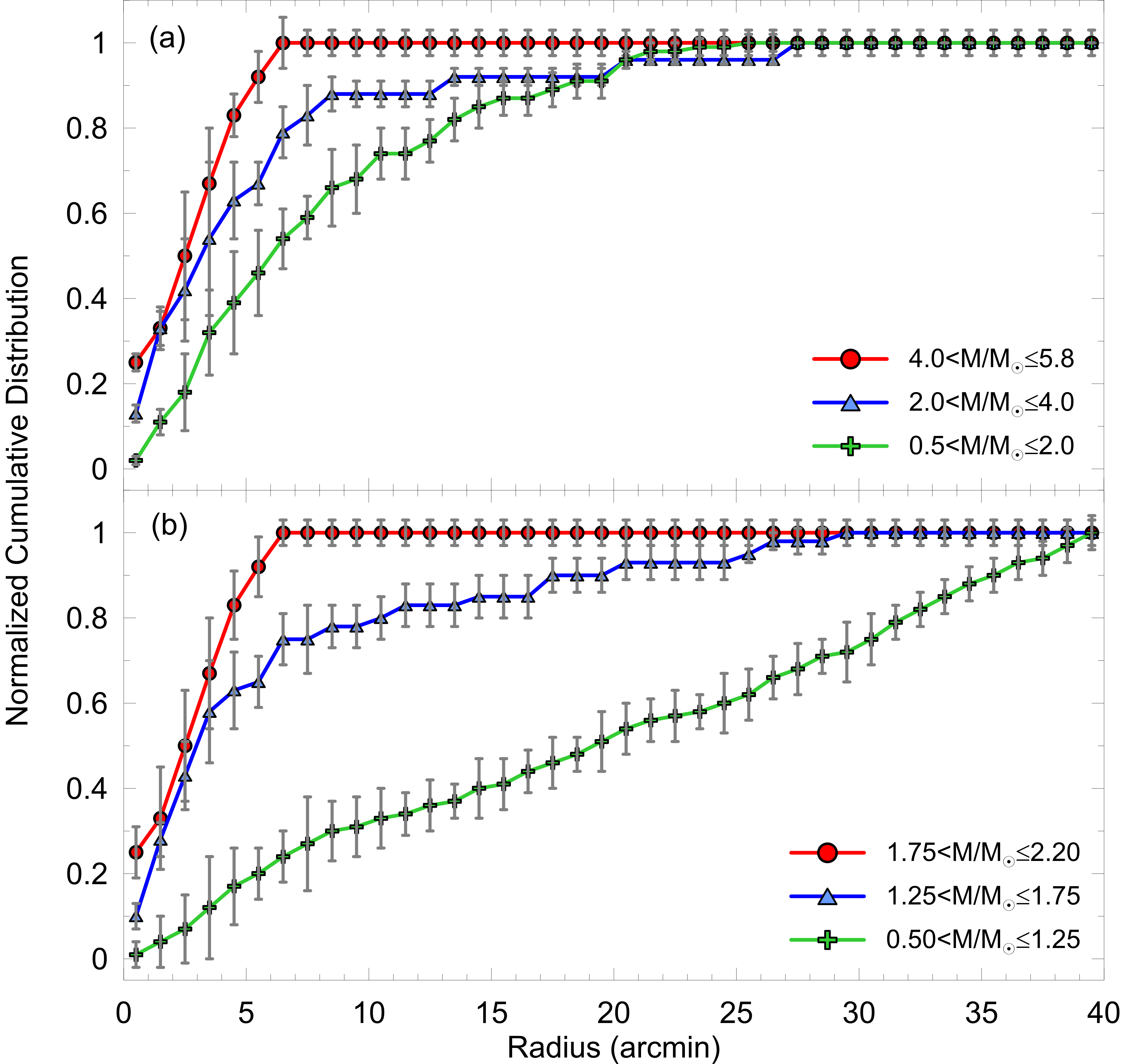}
\caption{\label{fig:radial_distributions}
The cumulative radial distribution of stars across various mass ranges for Roslund 3 (a) and Ruprecht 174 (b).}
\end {figure}

To assess mass segregation in the two clusters, we categorized the selected stars into three mass intervals with equal star counts. For Roslund 3, the mass ranges are $4 < M/M_{\odot} \leq 5.8$ (high-mass), $2 < M/M_{\odot} \leq 4$ (intermediate-mass), and $0.5 < M/M_{\odot} \leq 2$ (low-mass), while for Ruprecht 174 they are $1.75<M/M_{\odot} \leq 2.20$ (high-mass), $1.25 < M/M_{\odot} \leq 1.75$ (intermediate-mass), and $0.50 < M/M_{\odot} \leq 1.25$ (low-mass). The corresponding normalized cumulative radial distributions are presented in Figure~\ref{fig:radial_distributions}. The figure presents the normalized cumulative radial distributions of cluster member stars, stratified by stellar mass ranges. The analysis reveals a clear trend of mass segregation: high-mass stars are predominantly concentrated toward the central regions of the OCs, indicating their tendency to migrate inward, likely due to dynamical relaxation processes. In contrast, stars with intermediate and low masses exhibit relatively similar spatial distributions, predominantly occupying regions at larger radial distances from the cluster center in both OCs studied. This pattern supports the notion that mass-dependent spatial structuring is a significant feature of stellar population dynamics in OCs.

In order to assess the degree of mass segregation, we employed the Kolmogorov–Smirnov (K–S) test to statistically compare the radial distributions of high-mass stars against those of intermediate- and low-mass stellar populations. For the cluster Roslund 3, the test yielded confidence levels of 71\% for the high-mass group, 84\% for intermediate-mass stars, and 75\% for low-mass stars. Similarly, in Ruprecht 147, the corresponding values were 70\%, 85\%, and 74\%, respectively. These results, valid at a significance level of $P<0.05$, indicate that the null hypothesis, stating that the stellar subgroups originate from the same underlying distribution, can be rejected with moderate statistical confidence. This provides evidence that high-mass stars exhibit a distinct spatial distribution when compared to lower-mass stars, thereby supporting the presence of mass segregation in both OCs.

\begin{table*}
  \centering
  \renewcommand{\arraystretch}{1}
  \caption{Fundamental parameters of Roslund 3 and  Ruprecht 174.}
  \medskip
   {\small
        \begin{tabular}{lrr}
\hline
Parameter & Roslund 3 & Ruprecht 174 \\
\hline
($\alpha,~\delta)_{\rm J2000}$ (Sexagesimal)& 19:58:46.80, $+$20:30:32.5 & 20:43:25.44, $+37$:01:51.7 \\
($l, b)_{\rm J2000}$ (Decimal)              & 058.842, $-$04.684         & 78.012, $-03$.377          \\
$f_{0}$ (stars arcmin$^{-2}$)               & 26.832 $\pm$ 4.032         & 58.173 $\pm$ 3.189         \\
$f_{\rm bg}$ (stars arcmin$^{-2}$)          & 69.781 $\pm$ 0.383         & 41.335 $\pm$ 0.622         \\
$r_{\rm c}$ (arcmin)                        & 0.256 $\pm$ 0.055          & 0.336 $\pm$ 0.032          \\
$r_{\rm lim}$ (arcmin)                      & 2.5                        & 3.5                        \\
$r$ (pc)                                    & 1.23                       & 2.43                       \\
$C$                                         & 0.990 $\pm$ 0.093          & 1.018 $\pm$ 0.014          \\
Cluster members ($P\geq0.5$)                & 198                        & 397                        \\
$\mu_{\alpha}\cos \delta$ (mas yr$^{-1}$)   & $-0.401 \pm 0.003$         & $-3.139 \pm$ 0.006         \\
$\mu_{\delta}$ (mas yr$^{-1}$)              & $-5.169 \pm 0.003$         & $-4.729 \pm 0.007$         \\
$\varpi$ (mas)                              & 0.573 $\pm$ 0.004          & 0.412 $\pm$ 0.003          \\
$d_{\varpi}$ (pc)                           & 1745 $\pm$ 12              & 2427 $\pm$ 18              \\
$E(B-V)$ (mag)                              & 0.410 $\pm$ 0.046          & 0.615 $\pm$ 0.042          \\
$E(U-B)$ (mag)                              & 0.304 $\pm$ 0.033          & 0.462 $\pm$ 0.030          \\
$A_{\rm V}$ (mag)                           & 1.271 $\pm$ 0.143          & 1.907 $\pm$ 0.130          \\
$E({G_{\rm BP}-G_{\rm RP}})$ (mag)          & $0.478^{+0.15}_{-0.05}$    & $0.676^{+0.09}_{-0.14}$    \\
$A_{\rm G}$* (mag)                           & $0.89^{+0.28}_{-0.10}$     & $1.26^{+0.17}_{-0.26}$     \\
$[{\rm Fe/H}]$* (dex)                        & 0.030 $\pm$ 0.065          & 0.041 $\pm$ 0.064          \\
$Z$* (mass fraction)                         & 0.016                      & 0.017                      \\
($\mu_{V})_{\rm 0}$ (mag)                   & 12.407 $\pm$ 0.150         & 13.794 $\pm$ 0.144         \\
$d_{\rm iso}$* (pc)                          & 1687 $\pm$ 121             & 2385 $\pm$ 163             \\
$\tau$* (Myr)                                & 60 $\pm$ 6                 & 520 $\pm$ 50               \\
$(X, Y, Z)_{\odot}$ (pc)                    & (870, 1439, $138$)         & ($494$, 2329, $141$)       \\
$R_{\rm gc}$ (pc)                           & 7270 $\pm$ 24              & 7860 $\pm$ 15              \\
$\Gamma$                                    & 1.18 $\pm$ 0.13            & 1.53 $\pm$ 0.30            \\
$V_{\rm R}$ (km s$^{-1}$)                   & $-15.42 \pm 7.52$          & $-11.99 \pm 2.50$          \\
$U_{\rm LSR}$ (km s$^{-1}$)                 & 31.74 $\pm$ 6.07           & 68.86 $\pm$ 4.70           \\
$V_{\rm LSR}$ (km s$^{-1}$)                 & $-19.41 \pm 4.98$          & $-11.09\pm 1.57$           \\
$W_{\rm LSR}$ (km s$^{-1}$)                 & $-10.83 \pm 1.97$          & 2.25 $\pm$ 0.54            \\
$S_{_{\rm LSR}}$ (km s$^{-1}$)              & 38.75 $\pm$ 8.09           & 69.79 $\pm$ 4.98           \\
$R_{\rm a}$ (kpc)                           & 7.29 $\pm$ 0.05            & 7.99 $\pm$0.10             \\
$R_{\rm p}$ (kpc)                           & 5.95 $\pm$ 0.32            & 7.62 $\pm$ 0.02            \\
$Z_{\rm max}$ (pc)                          & 175 $\pm$ 26               & 128 $\pm$ 9                \\
$e$                                         & 0.101 $\pm$ 0.030          & 0.024 $\pm$ 0.006          \\
$P_{\rm orb}$ (Myr)                         & 182 $\pm$ 4                & 217 $\pm$ 4                \\
$R_{\rm teo}$ (kpc)                         & 5.95 $\pm$ 0.32            & 7.99 $\pm$ 0.10            \\
\hline
        \end{tabular}%
    }   \\
        * presents the results of the classical method adopted in the study.
    \label{tab:Final_table}%
\end{table*}%


\section{Summary and Discussion}

This study utilizes observational data from both ground-based CCD {\it UBV} photometry and space-based {\it Gaia} DR3 measurements to perform a comprehensive analysis, encompassing astrometric, astrophysical, kinematic, and dynamical aspects, of the OCs Roslund 3 and Ruprecht 174, located in the first Galactic quadrant. Two distinct approaches were employed to derive the clusters’ astrophysical parameters: the classical method and an MCMC-based technique, with particular attention given to addressing parameter degeneracies. The results reveal a strong consistency between the parameter estimates obtained from both methods, thereby confirming the robustness and reliability of the classical approach.

In the classical method, reddening and metallicity are individually estimated using {\it UBV} photometry through the application of TCDs. Keeping these parameters constant, we estimated distance and age of OCs to prevent possible parameter degeneracy. Meanwhile, the MCMC-based analysis of {\it Gaia} photometric data allows for the simultaneous determination of key astrophysical parameters, including absorption, distance, metallicity, and age. In the study we consider the distances from classical method to be the most reliable, as the method treats cluster reddening and metallicity as independent and fixed parameters during the analysis, while also incorporating membership information within a probabilistic framework. The main findings, along with a comparison to literature values (see Table~\ref{tab:literature}), are outlined below:

1)~The structural properties of the clusters were derived by fitting their RDPs with the \citet{King62} model. Limiting radii were set at $2'.5$ for Roslund 3 and $3'.5$ for Ruprecht 174, where field star contamination becomes significant. The concentration parameters, $C=0.990\pm 0.093$ and $C=1.018\pm 0.014$, indicate moderate central concentration, suggesting a moderate stage of dynamical evolution.

2)~This study combined astrometric and photometric methods to identify cluster members. Membership probabilities were calculated from {\it Gaia} DR3 data using the {\sc UPMASK} algorithm. Stars with $P \geq 0.5$ were plotted on CMDs and selected if they fell between the {\sc PARSEC} model main-sequence boundaries and were brighter than the completeness limit. This led to the identification of 198 members for Roslund 3 and 397 for Ruprecht 174. The {\it Gaia} and {\it UBV} catalogs were matched, applying similar criteria, resulting in 78 and 106 members for the two OCs, respectively.

3)~PM distributions and vectorial motions revealed high field star contamination for both OCs. The mean PM components $\langle \mu_{\alpha}\cos\delta, \mu_{\delta} \rangle$ were ($-0.401 \pm 0.003, -5.169 \pm 0.003$) mas yr$^{-1}$ for Roslund 3 and ($-3.139\pm 0.006, -4.729 \pm 0.007$) mas yr$^{-1}$ for Ruprecht 174, consistent with the VPD of membership-selected stars. The mean trigonometric parallaxes were estimated $0.573\pm 0.004$ and $0.412 \pm 0.003$ mas for Roslund 3 and Ruprecht 174, corresponding to distances of $1745 \pm 12$ and $2427 \pm 18$ pc, respectively. These values are consistent with recent \textit{Gaia}-based studies, such as \citet{Cantat-Gaudin18}, \citet{Dias21}, and \citet{Hunt23}. Earlier studies, particularly those predating \textit{Gaia}, report discrepant values with larger uncertainties \citep[e.g.][] {Kharchenko05, Loktin03}, likely due to the limitations in PM accuracy and member selection criteria.

4)~The reddening and metallicity of both OCs were estimated using TCDs. For Roslund 3, the derived values are $E(B-V) = 0.410 \pm 0.046$ mag and [Fe/H]=$0.030 \pm 0.065$ dex, while for Ruprecht 174 they are $E(B-V) = 0.615 \pm 0.042$ mag and [Fe/H] =$0.041 \pm 0.064$ dex. These reddening values lie within the range of those reported in the literature, \citep[e.g.][]{Kharchenko13, Dias14, Bonatto10}, although some older studies \citep[e.g.,][]{Turner93} report slightly lower values for Roslund 3. Metallicity estimates are sparse in the literature for both clusters; however, our values are broadly consistent with those found in recent studies such as \citet{Dias21} and \citet{Angelo21}, particularly when considering the error margins.

5)~Distances and ages of Roslund 3 and Ruprecht 174 were estimated using both $UBV$-based {\sc PARSEC} isochrones and {\it Gaia} DR3 data based MCMC analysis. $V$ vs $(U-B)$ and $V$ vs $(B-V)$ CMDs for stars with $P \geq 0.5$ yielded distance moduli of $12.407 \pm 0.150$ mag ($d_{\rm iso}=1687 \pm 121$ pc) for Roslund 3 and $13.794 \pm 0.144$ mag ($d_{\rm iso} = 2385 \pm 163$ pc) for Ruprecht 174. The corresponding ages were $60\pm 6$ Myr and $520\pm 50$ Myr, respectively. MCMC-derived posterior distributions of $A_{\rm G}$, $d$, $Z$, and $\log \tau$ were in agreement with the classical results, reinforcing the consistency between the two approaches. Literature values for distance and age vary considerably: Roslund 3 has been reported with distances ranging from 1080 pc \citep{Bossini19} to 2290 pc \citep{Turner93}, and ages between 41 Myr and 363 Myr \citep{Kharchenko05, Liu19}. Our findings fall within this broad range but suggest a more refined and internally consistent set of parameters. Similar age dispersion exists for Ruprecht 174, with literature values between 254 Myr \citep{Cantat-Gaudin_Anders20} and 800 Myr \citep{Bonatto10}. Our results, supported by both methods, place the clusters near the median of reported values.

6)~Mean radial velocities were derived from {\it Gaia} DR3 data based on identified member stars. For Roslund 3, the mean value was found to be $-15.42\pm 7.52$ km s$^{-1}$ from nine stars, whereas for Ruprecht 174, it was $-11.99\pm 2.50$ km s$^{-1}$ based on four stars. In comparison with literature, the radial velocity of both OCs has previously been reported as different values (see Table~\ref{tab:literature}). The discrepancies between our results and the literature values can be attributed to differences in the adopted \textit{Gaia} DR2 and \textit{Gaia} DR3, the sample size of member stars with available radial velocity measurements, and the membership selection criteria applied in each study.

7)~Galactic orbital analysis indicates that both OCs originated and remained within the solar circle. Their $V_{\rm LSR} > -50$ km s$^{-1}$ and $Z_{\rm max} < 200$ pc values confirm their membership in the thin-disk population.

8)~The PDMF slopes were found to be $\Gamma = 1.18 \pm 0.13$ for Roslund 3 and $\Gamma = 1.53 \pm 0.30$ for Ruprecht 174, both consistent within 1$\sigma$ of the \citet{Salpeter55} value.

9)~The estimated relaxation times for the OCs Roslund 3 and Ruprecht 174 were found to be approximately 24 and 263 Myr, respectively. These values suggest that both OCs have likely undergone sufficient dynamical evolution to reach a state of relaxation. In the case of Roslund 3, the reported value likely underestimates the true relaxation time due to observational incompleteness, particularly in the low-mass and binary star regimes. Therefore, it is plausible that the actual relaxation time exceeds the cluster's age, indicating that it has not yet achieved full dynamical relaxation.

\software{Astrometry.net \citep{Lang10}, Galpy \citep{Bovy15}, IRAF \citep{Tody86, Tody93},  PyRAF \citep{Science12}, SExtractor \citep{Bertin96}, Scipy  \citep{Scipy}, {\sc UPMASK} \citep{Krone-Martins14}}.
 

\begin{acknowledgments}
We thank the anonymous referee for their insightful and constructive suggestions, which significantly improved the paper. This study has been supported in part by the Scientific and Technological Research Council (T\"UB\.ITAK) 122F109. Funding was provided by the Scientific Research Projects Coordination Unit of Istanbul University as project number 40044. We thank T\"UB\.ITAK for partial support towards using the T100 telescope via project 18CT100-1396. We also thank the on-duty observers and members of the technical staff at the T\"UB\.ITAK National Observatory for their support before and during the observations. This study is a part of the MSc thesis of H\"ulya Karag\"oz.  We also made use of VizieR and Simbad databases at CDS, Strasbourg, France. We made use of data from the European Space Agency (ESA) mission \emph{Gaia}\footnote{https://www.cosmos.esa.int/gaia}, processed by the \emph{Gaia} Data Processing and Analysis Consortium (DPAC)\footnote{https://www.cosmos.esa.int/web/gaia/dpac/consortium}. Funding for DPAC has been provided by national institutions, in particular the institutions participating in the \emph{Gaia} Multilateral Agreement. We are grateful for the analysis system IRAF, which was distributed by the National Optical Astronomy Observatory (NOAO). NOAO was operated by the Association of Universities for Research in Astronomy (AURA) under a cooperative agreement with the National Science Foundation. PyRAF is a product of the Space Telescope Science Institute, which is operated by AURA for NASA. We thank the University of Queensland for collaboration software.
 
\end{acknowledgments}


\bibliography{open_clusters}{}
\bibliographystyle{aasjournal}



\end{document}